\documentclass[useAMS,usenatbib]{mn2e}
\usepackage{bethmacros}
\usepackage{epsfig}
\DeclareGraphicsExtensions{.eps, .jpg}

\title[Baryon Loss for Dark Satellites]{Mechanisms of Baryon Loss for Dark Satellites in Cosmological SPH Simulations}

\author[Nickerson et al.]{
{S. Nickerson$^{1,2}$, G. Stinson$^{1,3}$, H. M. P. Couchman$^{1}$, J. Bailin$^{1,4}$,  J. Wadsley$^{1}$}
\vspace*{6pt}\\
$^{1}$Department of Physics and Astronomy, McMaster University, Hamilton, Ontario, L8S 4M1, Canada \\
$^{2}$Institute for Theoretical Physics, University of Z\"{u}rich, Winterthurerstrasse 190, CH-8057, Switzerland \\
$^{3}$Jeremiah Horrocks Institute, University of Central Lancashire, Preston PR1 2HE, England \\
$^{4}$Astronomy Department, University of Michigan, 830 Dennison Bldg., 500 Church St., Ann Arbor, MI 48109-1042, United States}

\begin{document}
\maketitle
\label{firstpage}

\begin{abstract}
We present a study of satellites in orbit around a high-resolution, smoothed particle hydrodynamics (SPH) galaxy simulated in a cosmological context.  The simulated galaxy is approximately the same mass as the Milky Way.  The cumulative number of luminous satellites at $z=0$ is similar to the observed system of satellites orbiting the Milky Way although an analysis of the satellite mass function reveals an order of magnitude more dark satellites than luminous. Some of the dark subhalos are more massive than some of the luminous subhalos at $z=0$.  What separates luminous and dark subhalos is not their mass at $z=0$, but the maximum mass the subhalos ever achieve. We study the effect of four mass-loss mechanisms on the subhalos:  ultraviolet (UV) ionising radiation, ram pressure stripping, tidal stripping, and stellar feedback, and compare the impact of each of these four mechanisms on the satellites. In the lowest mass subhalos, UV is responsible for the majority of the baryonic mass loss.  Ram pressure stripping removes whatever mass remains from the low mass satellites.  More massive subhalos have deeper potential wells and retain more mass during reionisation.  However, as satellites pass near the centre of the main halo, tidal forces cause significant mass loss from satellites of all masses.  Satellites that are tidally stripped from the outside can account for the luminous satellites that are lower mass than some of the dark satellites.  Stellar feedback has the greatest impact on medium mass satellites that had formed stars, but lost all their gas by $z=0$. Our results demonstrate that the missing satellite problem is not an intractable issue with the cold dark matter cosmology, but is rather a manifestation of baryonic processes.
\end{abstract}

\begin{keywords}
galaxies: dwarf --- cosmology: theory --- galaxies: evolution --- methods: N-Body simulations --- methods: numerical
\end{keywords}

\section{Introduction}
\label{sec:intro}

The $\Lambda$ Cold Dark Matter ($\Lambda$CDM) cosmology is currently the most widely accepted and successful paradigm for describing the Universe \citep{Blumenthal1984,Davis1985, Gramann1988,Peebles2003}. Consequently, structure in the Universe forms hierarchically, as shown analytically and in simulations \citep[e.g.][]{PressSchechter1974,WhiteRees1978,Davis1985}. First, dark matter collapses into small halos, and later these collect as subhalos into galaxies, where gas cools into a disk to form stars. In especially dense regions of the Universe galaxies bind together gravitationally into galaxy clusters.

In a hierarchical Universe, substructure is expected to be invariant at all scales of interest. In some of the earliest simulations that were able to resolve substructures, \citet{Moore1999} found that dark matter-only simulations of galaxies and galaxy clusters had the same number of substructures relative to the total mass of the system.  A comparison of the simulations to observations showed that the simulated galaxy cluster matched the quantity of substructure in the nearby Virgo cluster, but that the simulated galaxy had significantly more substructure than the Local Group. Using constrained simulations of a system similar to the Local Group, \citet{Klypin1999} found far more substructure than what has been observed.  This discrepancy between $\Lambda$CDM and observations is known as the ``missing satellites problem''.  \citet{Kravtsov2010} provides a recent review of the progress made towards solving this problem.

Large observational surveys have also discovered a new class of ultra-faint galaxies \citep{Willman2005,Belokurov2007, Koposov2008}.  The detection of such galaxies slightly lessens the number of satellites that are ``missing'' from the Local Group. Early observations of their velocity dispersions \citep{Simon2007} show that stars may form in halos of lower mass than previously believed possible, although the typical star formation efficiency in these low mass halos must be extremely low given that the halo mass function rises steeply at these masses \citep{Tollerud2009,Guo2010b}. A key question is whether there is a minimum mass halo in which stars can form, and if there is significant scatter in the star formation efficiency at a given halo mass. The low mean star formation efficiency at these masses might be driven by halo-to-halo variation, or a steep, universal relation between star formation efficiency and halo mass. The ultra faint satellites have been detected down to $M_{V}\approx-2$, which corresponds to 100 $L_{\sun}$, below the resolution of the simulations studied here. It is impossible to give a full census of such objects from the simulations.  However, we show that some small objects do form in the simulations.

Generally, there are two paths pursued to solve the missing satellites problem. One is to alter the cosmological paradigm.  Examples of this include self-interacting dark matter in which subhalos are destroyed through self-annihilation \citep{Spergel2000},  initially warm dark matter out of which small structures do not form \citep{Dalcanton2001}, or removing small scale perturbations from the primordial power spectrum \citep{Zentner2003}.  Recent gravitational lensing studies have discovered dark substructures \citep{Dalal2002,Mao2004}, so it appears $\Lambda$CDM is consistent with observations and we must find what physical mechanisms play the largest role in darkening small galactic halos.  This leads to the other path, which is to consider the effects of baryonic physics, such as stellar feedback \citep{Dekel1986,MacLow1999} and UV ionisation \citep{Efstathiou1992,Quinn1996,Bullock2000}, which might render many satellites dark.

The four primary mechanisms that can remove mass from halos are:
\begin{itemize}
\item {\bf UV ionisation:}  luminous objects emit UV radiation that ionises hydrogen and sets a background temperature above the virial temperature of the subhalo.
\item {\bf Ram pressure stripping: }  as a satellite passes through the hot halo gas, the incident gas pressure becomes stronger than the gravitational force of the satellite and gas is thus removed.
\item {\bf Stellar feedback:} stellar winds and supernovae inject sufficient energy into the interstellar medium (ISM) of the small galaxy so that some or all of the ISM is ejected.
\item {\bf Tidal stripping:} as a satellite orbits close to a larger host galaxy, the tidal forces become sufficient to remove material. Unlike the other three mechanisms mentioned above this is the only one that can remove collisionless matter, namely dark matter and stars, as well as gas from a subhalo.
\end{itemize}

Previous efforts have been made at examining these mechanisms in detail.  Early efforts were analytical due to the large dynamic range necessary to properly simulate substructures, but recent simulations have allowed a closer look at satellites.  \citet{Dekel2003} compared careful observations of many dwarfs with an analytical model based on the effect of supernova feedback and found that supernova feedback defines the line between low and high luminosity dwarf galaxies.  \citet{Kravtsov2004} used high resolution, cosmological simulations to study the role of tidal stripping in the mass evolution of satellites.  They concluded that the combined effect of tides and ionisation could produce a Milky Way-like satellite luminosity function. \citet{Read2006} considered both supernovae-driven winds and ionisation in cosmological simulations and found that ionisation was critical to make their simulations agree with observed luminosity functions.  \citet{Governato2007} found in another series of cosmological simulations that UV background dramatically reduced the number of luminous subhalos, but that stellar feedback was required to make the simulated luminosity functions the same as those observed.  More recent cosmological simulations by \citet{Okamoto2010} have varied the strength of a kinetic supernova wind feedback to determine exactly how much energy is required to produce the observed luminosity function. \citet{Klimentowski2010} saw how tidal stripping determines a subhalo's baryon content and final morphology. \citet{Wadepuhl2011} introduced black holes into their simulations and found that the black holes are not massive enough in subhalos to have an effect on their luminosities. They did, however, find that wind-driven galactic outflows can reduce the number of high mass satellites, and cosmic rays can suppress the luminosity of low mass subhalos. Each of these models successfully fit the data by studying in detail one or two mechanisms, while we will consider all four within SPH simulations.

Recent semi-analytic models also show some success at reproducing the observed satellite luminosity function.  In these, only the more massive subhalos (\citet{OkomotoFrenk2009}, \citet{Guo2010a}, \citet{Maccio2009}) retain stars.  One semi-analytic model of an N-body simulation of a Milky Way-like halo \citep{Li2009} reveals luminous subhalos whose mass in dark matter spanned one order of magnitude, while the luminosity ranged over five orders of magnitude, matching observations. There were also many more dark matter-only subhalos present, whose mass spanned three orders of magnitude.

\citet{Mayer2006} pointed out the importance of the combined effects of each mechanism. They simulated individual satellites falling into a static gravitational potential filled with hot, dense gas.  In these simulations, tidal forces excite star formation and thus stellar feedback, as well as reshaping the gas distribution so that it can be more easily stripped due to ram pressure.  \citet{Mayer2006} called this combination of processes ``tidal stirring'' and found that it can remove enough gas from dwarf irregulars to turn them into gas-free dwarf spheroidals.

The purpose here is to discover which mechanisms tore the baryons off the subhalos, focusing on UV background, tidal stripping, ram pressure stripping, and stellar feedback. For the first time, we will explicitly track the causes behind the departure of individual gas particles from their subhalo in order to construct a comprehensive picture showing the relative strength of each gas loss mechanism.  We only analyse the satellites inside the virial radius, so we expect the satellites we are analysing to be similar to dwarf spheroidal (dSph) galaxies, a population that dominates the satellite population of the Milky Way within $r_{vir}$. Dwarf spheroidals are gas-poor \citep[e.g. Fig 3 in][]{Grebel2003} containing as little as $10^{4}$M$_{\sun}$ to undetectable amounts, but generally they continued forming stars until recently \citep{Skillman2007}.

\S\ref{sec:method} establishes the background behind the tools we used in this work. \S\ref{sec:simlumefns} introduces the subhalo population of our host galaxy g15784 at $z=0$, while \S\ref{sec:histhalos} details the history of these subhalos and how we determine the causes of baryon loss. The concluding \S\ref{sec:conclu} discusses the implications of our findings and areas for future work.

\section{Method}
\label{sec:method}

We analyse the evolution of the subhalos of two galaxies (g15784 and g5664) from the McMaster Unbiased Galaxy Simulations (MUGS).  The purpose of MUGS is to provide a sample of M$^*$ galaxies simulated using SPH at high resolution.  A full description of MUGS can be found in \citet{Stinson2010}, but we briefly summarize it here.  The MUGS sample is chosen from a 50 $h^{-1}$ Mpc volume of a WMAP3 $\Lambda$CDM universe ($H_{0}$=73 km s$^{-1}$ Mpc$^{-1}$, $\Omega_{m}$=0.24, $\Omega_{\Lambda}$=0.76, $\Omega_{baryon}$=0.04, and $\sigma_{8}$=0.79) \citep{Spergel2007}. It consists of a random selection from the galaxies with halo masses between $\approx5\times10^{11}$ M$_{\sun}$ and $\approx2\times10^{12}$~M$_{\sun}$ that did not evolve near to structures more massive than $5.0\times10^{11}$M$_{\sun}$ within 2.7 Mpc. While this would have eliminated the Milky Way from our sample, there is no evidence for a past interaction between the Milky Way and M31 and so the Milky Way's satellite population should be unaffected by its near neighbor. The sample is unbiased with regards to angular momentum, merger history, and less massive neighbors and it is hoped that the sample will reproduce the observed spread in galaxy properties. 

The only bias is random. The selected galaxies are simulated with the commonly-used zoom technique that focuses resolution on individual galaxies while maintaining the large scale torques necessary to give galaxies their angular momentum. The initial dark matter, gas and star particle masses are $1.1\times 10^{6}$~M$_{\sun}$, $2.2 \times 10^{5}$~M$_{\sun}$ and $6.3 \times 10^{4}$~M$_{\sun}$ respectively. Each type of particle uses a constant gravitational softening length of 310 pc. Most of this paper will be dedicated to the subhalos of one of these galaxies, g15784, which has a mass of $1.43\times10^{12}$~M$_{\sun}$. Its disk has a mass of $3.27\times10^{10}$~M$_{\sun}$ and the bulge a mass of $5.49\times10^{10}$~M$_{\sun}$, based on kinematic decomposition \citep{Stinson2010}. We also utilize the $5.2\times10^{11}$~M$_{\sun}$ galaxy g5664 to see how the luminosity function changes depending on the presence of stellar feedback and UV background.

Outputs were at most 214 Myr apart, especially at lower redshift, with irregular outputs at key times. Outputs were much closer together at high redshifts, typically 107 Myr apart.

MUGS was run using the SPH code \textsc{gasoline} \citep{Wadsley2004}.  \textsc{gasoline} includes low temperature metal cooling (described in \citet{Shen2010} and briefly here), UV background radiation, star formation, and physically-motivated stellar feedback.  The metal cooling grid is constructed using CLOUDY (version 07.02, last described by \citet{Ferland1998}), assuming ionisation equilibrium. A uniform ultraviolet ionising background, adopted from Haardt \& Madau (in preperation; see \citet{HaardtMadau}), is used in order to calculate the metal cooling rates self-consistently. It starts to have an effect at $z=9.9$.

\subsection{Star Formation and Feedback}

The star formation and feedback recipes are the ``blastwave model" described in detail in \citet{Stinson2006}.  They are summarized as follows. Gas particles must be dense ($n_{\rm min}=1.0~\mathrm{cm}^{-3}$) and cool ($T_{\rm max}$ = 15,000 K) to form stars. A subset of the particles that pass these criteria are randomly selected to form stars based on the commonly used star formation equation,
\begin{equation}
\frac{dM_{\star}}{dt} = c^{\star} \frac{M_{gas}}{t_{dyn}}
\end{equation}
where $M_{\star}$ is mass of stars created, $c^{\star}$ is a constant star formation efficiency factor, $M_{gas}$ is the mass of the gas particle spawning the star, $dt$ is how often star formation is calculated (1 Myr in all of the simulations described in this paper) and $t_{dyn}$ is the gas dynamical time. The constant parameter, $c^{\star}$, is tuned to 0.05 so that the simulated isolated Model Milky Way used in \citet{Stinson2006} matches the \citet{Kenn1998} Schmidt Law, and then $c^\star$ is left fixed for all subsequent applications of the code.

At the resolution of these simulations, each star particle represents a large group of stars (6.32 $\times 10^4$ M$_{\sun}$). Thus, each particle has its stars partitioned into mass bins based on the initial mass function presented in \citet{Kroupa1993}. These masses are correlated to stellar lifetimes as described in \citet{Raiteri1996}. Stars larger than $8~$M$_{\sun}$ explode as supernovae during the timestep that overlaps their stellar lifetime after their birth time. The explosion of these stars is treated using the analytic model for blastwaves presented in \citet{MO77} as described in detail in \citet{Stinson2006}. While the blast radius is calculated using the full energy output of the supernova, less than half of that energy is transferred to the surrounding ISM, $E_{SN}=4\times10^{50}$ ergs. The rest of the supernova energy is assumed to be radiated away.

To capture the behavior of clustered star formation, we stochastically determine when a star particle releases feedback energy,

\begin{eqnarray}
p & = & \frac{N_{SNII}~mod~N_{SNQ}}{N_{SNQ}} \\
N_{ESN} & = &\left\lfloor\frac{N_{SNII}}{N_{SNQ}}\right\rfloor + \left\{
\begin{array}{l l}
0 & \quad ,r \leq p \\
N_{SNQ} & \quad ,r > p \\
\end{array} \right.
\end{eqnarray}
where $N_{SNII}$ is the number of supernovae calculated to explode during that star formation timestep, $N_{SNQ}$ is the ``supernova quantum'', i.e. the number of supernova required per explosion (fixed at $30$, the number of supernovae expected from the star particles in our simulation), and $N_{ESN}$ is the total number of supernova explosions that will have their energy distributed during a the star formation timestep. If the probability, $p$, is greater than a random number, $r$, selected between 0 and 1, $N_{SNQ}$ supernovae's worth of energy is released. This causes SN energy to be released in quantized packets over the 35 Myr until the largest star remaining is $< 8$ M$_{\sun}$.

\subsection{Group Finding: Amiga Halo Finder}

In order to identify the host galaxy and its subhalos, we used the Amiga Halo Finder (AHF) \citep{Knollmann2009}. AHF is based on the spherical overdensity method for finding halos.  It is able to identify density peaks using an adaptive mesh algorithm. Once the density peaks are identified, AHF cuts out halos (and subhalos) using isodensity contours.  Particles belonging to subhalos are distinguished from those of the background halo using a simple unbinding procedure to determine whether the particles are gravitationally bound to the subhalo.  We base our analysis on a minimum group size of 50 particles, which is $2.2\times10^{7}$~M$_{\sun}$ when the group only contains dark matter but could be less massive if it also contains gas and star particles.  Our analysis is restricted to those satellites identified by AHF as lying inside the virial radius ($r_{vir}$) of the halo.  For g15784, $r_{vir}=240$~kpc, while for g5664, $r_{vir}=152$~kpc.

\subsection{Merger Trees}
\label{sec:mt}
We traced the histories of each subhalo in the galaxy. First, we identified groups at every output 100 Myr apart with AHF and then traced the particles present in the subhalos at $z=0$ back through the simulation including any gas out of which stars formed.  For the sake of clarity, let us call the subhalo of interest ``Alpha''. At each output, we note every group that contains Alpha's particles. The group that had the largest number of Alpha's particles is set as Alpha's progenitor at that output. In this way, we trace the properties of each subhalo through time, including mass, distance from the host galaxy, and temperature.

Because this method only depends on the number of particles, a subhalo can ``jump'' in position space between outputs, switching between subhalos with different groups of particles. Out of the higher mass subhalos, such behaviour was only observed in three of them. This does not affect our results because the jumps only occurred at high redshifts and between subhalos of comparable mass and position that were soon to merge. The pivotal point of analysis in each subhalo's history is its maximum mass, as will 
be show, in \S \ref{sec:histhalos}, and it occurs well after any jumping behavior we see. Also, we will only look at the \textit{last} time that a particle leaves its subhalo in order to prevent a single gas particle from being double-counted. Therefore, any event of gas being ``lost'' due to subhalo jumping will automatically be removed from the analysis.

\section{Simulated Luminosity Functions}
\label{sec:simlumefns}
The first analysis we undertook was to compare the satellite luminosity function of g15784 to observations. We found that the cumulative number is similar to the Milky Way's, though there was an excess of high luminosity satellites and so the shape of the cumulative functions did not match. We also re-simulated a smaller galaxy, g5664, with and without the UV background and stellar feedback to compare the effects the presence these two mechanisms have on the subhalo population as a whole.

\subsection{The Main Halo: g15784}
\label{sec:g15784}
AHF found 107 satellites inside $r_{vir}$ of g15784.  In order to determine the luminosity of each subhalo, we treated each star particle as its own stellar population, where MUGS allowed us to model a population of stars with a distribution of ages.  We based the brightness of the stars on the luminosity grid provided by CMD 2.1 \citep{Leitherer1999,Marigo2008}.  Using the grid, we performed a bilinear interpolation over the stellar ages and metallicities of each star particle and then summed the luminosities of all the star particles in each satellite to derive a stellar magnitude for the satellites.  We neglect the effects of dust extinction since dwarf galaxies are low metallicity and rarely appear dust obscured \citep{Mateo1998,Lisenfeld1998}. This MUGS galaxy has an effective resolution of $2048^3$, and we will also compare g15784 to a lower resolution run with an effective resolution of $1024^3$.

\begin{figure}
\centering
\resizebox{0.45\textwidth}{!}{\includegraphics{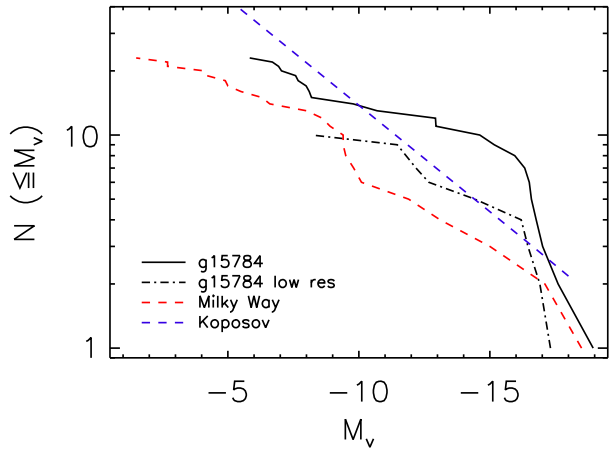}}
\caption{The cumulative V-band luminosity function of the host galaxy's subhalos at $z=0$, in solid black. Observational data of the Milky Way as gathered by \citet{Tollerud2008} is shown in dashed red, while a theoretical completion-corrected function from \citet{Koposov2009} is shown in dashed blue. The black dash-dotted line is the luminosity function from a lower resolution run of the same galaxy where the initial gas particle mass is $\approx 10^{6}$ M$_{\sun}$.} (Note that the observed ultra-faint dwarfs extend to a much lower luminosity). Within the resolution limit, all three functions lie within an order magnitude of each other.
\label{fig:cumemag}
\end{figure}

Figure \ref{fig:cumemag} shows the cumulative luminosity function of the host galaxy's subhalo population in the V-band at $z=0$.  The results show that down to $M_{V} \approx -6$, the dimmest subhalo in our simulation, the number of satellites is 23 compared with 20 to 21 for the Milky Way at a similar magnitude. Thus, when all the relevant baryonic processes are included the order-of-magnitude missing satellites problem \citep{Moore1999, Klypin1999} disappears as has been seen in other simulations of comparable resolution \citep{Okamoto2010, Knebe2010, Wadepuhl2011}. 

For comparison to observations we include two lines for reference. This first is a recent compilation of the classical satellites and the new ultra-faint dwarf galaxies from \citet{Tollerud2008}. The census of ultra-faint dwarfs is certainly incomplete, both due to the limited sky coverage of the Sloan Digital Sky Survey and the difficulty in detecting very faint galaxies. These galaxies extend to lower luminosities than our simulations can resolve, where our faintest satellite has one star particle. The resolution of our simulations is not sufficient to make robust predictions regarding the new classes of ultra-faint dwarfs. 

The second is \citet{Koposov2008} who modelled these effects and derived a corrected luminosity function, $\frac{dN}{dM_{V}}=10\times10^{0.1(M_{V}+5)}$ for $-18 \le M_{V} \le -2$, that would represent a theoretically complete set of satellite galaxies; we have also plotted this function in Figure~\ref{fig:cumemag}. Our host galaxy g15784 has a mass close to the Milky Way's, so it is reasonable that its cumulative number of luminous subhalos is comparable to the Milky Way's, as indeed it is. In other words, \emph{our simulated galaxy does not suffer from the missing satellite problem}. The major difference between the simulations and observations is an excess of brighter satellites in the simulations, and that our star formation recipe may form too many stars. This causes an extra ``knee'' in the shape of our luminosity function.

The dash-dotted line in Figure \ref{fig:cumemag} shows the luminosity function of a lower resolution simulation exactly the same as g15784, but with half the spatial resolution and an initial gas particle mass of $\approx 10^{6}$ M$_{\sun}$. From this it is clear that decreasing resolution decreases the number and luminosity of subhalos. The lowest luminosity subhalo, at $M_V \approx -8.2$, contains a single star; in our higher resolution run this magnitude corresponds to 10 star particles. We discuss resolution effects in \S \ref{sec:res}.

\begin{figure}
\centering
\resizebox{0.45\textwidth}{!}{\includegraphics{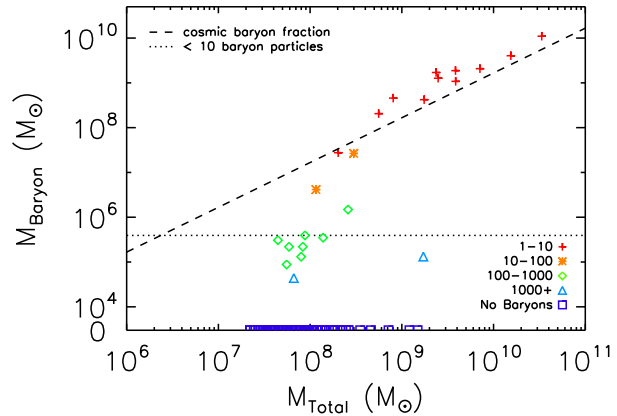}}
\caption{Baryon mass versus total mass for our subhalo population. The symbols correspond to mass to light ratio (total mass divided by baryonic mass), and the dashed  black line is where subhalos with the cosmic baryon fraction would lie. The lower bound of the mass-to-light ratio is inclusive. The horizontal dotted line is the most luminous of the satellites with fewer than $10$ baryon particles, corresponding to our resolution limit. Most striking here is that the luminous subhalos, the brightest of which follow the power law, and dark subhalos overlap in total mass.}
\label{fig:msmtot}
\end{figure}

Figure \ref{fig:msmtot} shows the baryonic mass of each subhalo as a function of total mass at $z=0$.  The dashed line shows where subhalos that obey the cosmic baryon fraction would lie. The subhalos contain systematically fewer baryons below $5\times10^8$ M$_{\sun}$. This falloff is similar to the low mass dropoff found by \citet{McGaugh2010} in the observed baryonic Tully-Fisher relationship.  In these lower mass halos, baryons are preferentially stripped.  As we shall see, these halos have also lost a significant amount of dark mass. However, almost all the lower mass halos also have fewer than $10$ baryons.

Additionally, below $2\times10^9$ M$_{\sun}$ there are many halos that contain no baryon particles according to our simulation's resolution at all and are thus dark. Of the 23 satellites that contain baryons, only 10 contain gas with the maximum gas fraction being 4 \% of the total mass. This fraction might seem low if compared to high gas fractions found in isolated dwarf irregular galaxies \citep{Geha2006}, but the subhalos considered here are all within $r_{vir}$ and are more appropriately compared with the dwarf spheroidals presented in \citet{Grebel2003}, which contain little gas. We must stress that when we say that a subhalo ``has no baryons'' we mean in the sense of our simulation's resolution there are no baryon particles. Actual dark subhalos will always have at least some trace quantities of gas.

\begin{figure}
\centering
\resizebox{0.45\textwidth}{!}{\includegraphics{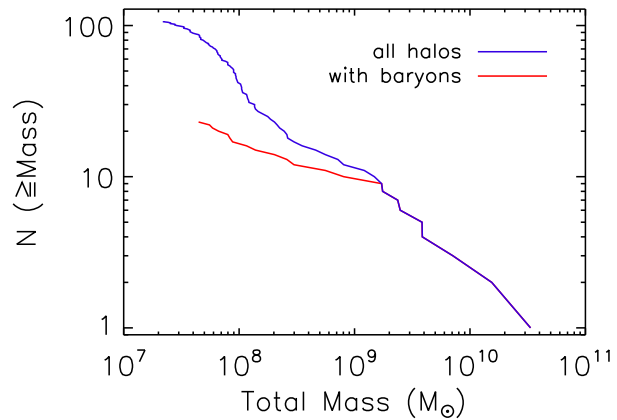}}
\caption{The cumulative mass function of our host galaxy's subhalos at $z=0$, divided into all subhalos in blue, and those with baryons in red. The total number of subhalos is over a hundred, but the total number containing baryons is almost an order of magnitude fewer.}
\label{fig:cumemass}
\end{figure}

Figure \ref{fig:cumemass} shows the cumulative mass function at $z=0$ for all the satellites as well as the subset that formed stars. The total satellite mass function is similar to the collisionless, dark matter-only simulations of \citet{Moore1999} and \citet{Klypin1999} while the mass function of luminous satellites is closer to what is observed. There is about an order of magnitude more subhalos in total than those that contain baryons. How these dark subhalos lost their gas and their stars, if they ever had any, is the key to understanding the missing satellites problem. We will investigate gas removal mechanisms, the reason some subhalos are light and some are dark, and why the only 23 subhalos that are luminous are not also the 23 most massive in \S \ref{sec:histhalos}.

\subsection{Effects of Stellar Feedback and Ultraviolet Background: g5664}
\label{sec:g5664}

One test to delve deeper into how the cumulative mass function compares with dark matter-only simulations while the cumulative number of luminous satellites is consistent with observations is to re-simulate a galaxy with and without the UV and stellar feedback. We re-simulated a second, smaller host galaxy, g5664, with a mass of $5.2\times10^{11}$ M$_{\sun}$, three times with different baryonic physics:
\begin{enumerate}
\item[(a)] with the standard MUGS simulation including UV and stellar feedback,
\item[(b)] with UV but no stellar feedback, and
\item[(c)] with neither UV nor stellar feedback.
\end{enumerate}
This galaxy contained half the number of gas and star particles as g15784 and thus was faster to run, which is why it was chosen for the parameter comparison. Since g15784 is more massive and has a larger $r_{vir}$, it will contain more and likely more massive satellites.  Since the UV background and stellar feedback more strongly effect low mass galaxies, g15784 will contain more luminous satellites. However, this should not affect the relative comparison of the different resimulations of g5664. The relationship between these mechanisms and the mass of the satellites will be explored in \S \ref{sec:histhalos}, in addition to how these mechanisms affect the satellites throughout their history.

\begin{figure}
\centering
\resizebox{0.45\textwidth}{!}{\includegraphics{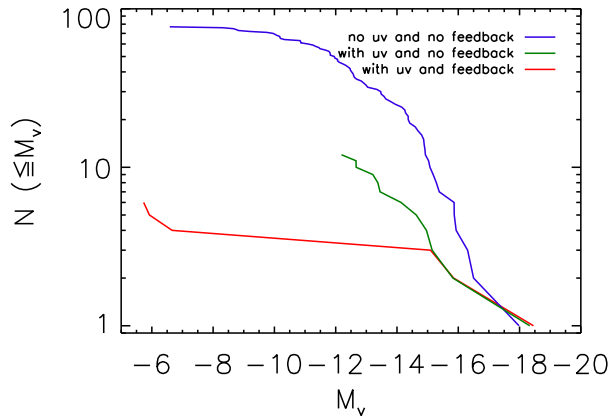}}
\caption{Cumulative luminosity function for g5664, comparing runs with three different conditions on the baryonic physics. The run with UV background and stellar feedback is in red (a). The run with UV background, but no feedback is in green (b). The run with neither UV background nor stellar feedback is in blue (c).  Both feedback and the UV background reduce the total number of luminous subhalos.}
\label{fig:g5664mag}
\end{figure}

Figure \ref{fig:g5664mag} shows the cumulative luminosity function for the three simulations at $z=0$. It confirms the reduction in number of satellites.  There about one-fourth as many in g5664 as in g15784. Nearly every satellite in the simulation run without feedback and UV (c) contains stars.  Conversely, many fewer satellites contain stars in the simulation that includes feedback (a).  It is apparent that the UV ionisation plays a large role in stopping stars from forming in many subhalos.  When stellar feedback is added, there is little effect on subhalos brighter than $M_{V}\approx-15$, but fainter subhalos are only populated with a few stars whose feedback was effective at eliminating star formation for the rest of the simulation. The $M_{V}\approx-15$ threshold is similar to the flattening seen in Figure \ref{fig:cumemag} for both the simulated g15784 and the observed Local Group mass function.

The UV-only simulation (b) contains several medium luminosity subhalos but none that are extremely faint, stopping at $M_{V}\approx-12$. It is curious then that simulation (a), with both feedback and UV, lacks satellites between $-8 < M_{V} < -15$, but contains two very faint satellites at $M_{V} \approx -6$. The two luminosity functions diverge at  $M_{V} \approx -15$, which corresponds to about $10^4$ star particles in all three runs of g5664.  The origin of this situation is unclear from inspection of the luminosity functions alone. In \S \ref{sec:histhalos}, however, when we track the mechanisms of baryon loss through time an explanation for this phenomenon will arise showing that stellar feedback preferentially strips medium mass satellites.

\subsection{Resolution}
\label{sec:res}

Satellite galaxies are the most difficult objects to resolve in simulations, and so they show the strongest resolution effects. Figure \ref{fig:cumemag} begins to show how resolution can affect satellites. It is harder to detect low mass satellites in lower resolution simulations, and effects such as tides and ram pressure stripping are more accurately captured as resolution increases.

Other authors have discussed the resolution effects on their satellite luminosity functions in simulations similar to ours. \cite{Libeskind2007} ran simulations at a lower resolution than ours, studying the satellite systems of several $\Lambda$CDM galaxies with their gas resolved to $\approx10^{6}$ M$_{\sun}$. They compared their luminosity functions to semi-analytic models of resolution over four different orders of magnitude and found a convergence at $M_{V}\approx-12$. \cite{Okamoto2010} studied the effects of several feedback models on the satellite populations on host galaxies around the same mass as our g15784 and with similar gas particle masses. Matching the circular velocity of their satellites to a power law, they concluded that their satellites were well resolved down to satellites with at least 10 star particles. \cite{Wadepuhl2011} compared a low resolution simulation (whose gas particle mass is $\approx10^{5}$ M$_{\sun}$ corresponding to our high resolution simulation) to their high resolution simulation with gas mass of $\approx10^{4}$ M$_{\sun}$ and found them convergent up to $M_{V}\approx-8$. 

We compared our high resolution simulation of g15784, with initial gas particle mass $\approx 10^5$ M$_{\sun}$, to a low resolution simulation, with initial gas particle mass $\approx 10^6$ M$_{\sun}$, and found that their luminosity functions were similar in their region of overlap down to $M_{V} \approx -8.2$. The major difference is that at higher resolution there appears to be a slight overabundance of highly luminous subhalos. \citet{Christensen2010} examined the effects of resolution on galaxies ranging from $10^{13}$ to $10^{9}$ M$_{\sun}$ and found that $10^4$ gas particles at each galaxy's maximum mass are required before the star formation recipe used here converges.

Subhalos that formed stars in our high resolution g15784 simulation had their star formation drop off at about $10^4$ star particles at $z=0$, corresponding to the number of gas particles being at least $\approx 5\times10^3$ at maximum mass. The star count corresponds to $M_{V} \approx -15$ in our high resolution simulation. Based on \citet{Christensen2010}, we expect to see a decrease in star formation across this range. However, it is difficult to conclude that the reduction in star formation is solely due to resolution since it corresponds to the mass range between $10^9$ and $5\times10^{10}$ M$_{\sun}$ where feedback has its strongest effects. Our low resolution simulation of g15784 only has one subhalo at $M_{V} \approx -17$ with $\approx 7\times10^3$ star particles, and only two more around $M_{V} \approx -16$ with $\approx 10^3$ star particles. The discrepancy between high and low resolutions is explained by the lack of particles in the low resolution run.

Some of our subhalos contain many fewer particles than \cite{Christensen2010} suggest is necessary to resolve star formation, but we include them for completeness. We will still show and analyze all luminous and dark subhalos with comprehensive histories (as explained early in \S \ref{sec:histhalos}). With higher resolution, \cite{Christensen2010} suggest that these satellites would contain slightly more stars since gas could form a disk and become denser than the star formation threshold. However, we note that it is hard to draw conclusions without the higher resolution simulations because of the non-linear interaction of star formation and stellar feedback, particularly considering that the present simulation over-predicts the number of highly luminous satellites (Stinson et al. 2010).

\section{Gas Loss Mechanisms}
\label{sec:histhalos}
The previous section showed that both stellar and UV feedback play a significant role in the evolution of satellite galaxies. In order to gain an overall picture of how stellar and UV feedback work with ram pressure and tidal stripping, we will focus on g15784 and study the detailed, particle-by-particle mass loss history of its subhalos.

We traced the evolution of 85 of the 107 subhalos identified by AHF in g15784.  We were unable to perform a detailed trace of every subhalo for two reasons: 
\begin{enumerate}
 \item[(1)] 17 low mass subhalos did not maintain 50 member particles throughout the simulation, which is $2.2\times10^{7}$M$_{\sun}$ when the group only contains dark matter but could be less massive if it also contains gas and star particles. 
  \item[(2)] 5 halos were spatially coincident with another subhalo during one output, and thus appeared to gain a large amount of mass.  Since our analysis focused on  cumulative baryon loss and the subhalos' maximum mass these sudden spikes in mass would invalidate any results including those halos.
\end{enumerate}

10 of the 85 subhalos we analysed contained both gas and stars at $z=0$, with total masses ranging from $5.6\times10^{8}$~M$_{\sun}$ to $3.3\times10^{10}$~M$_{\sun}$. Of the subhalos that did not have gas at $z=0$, 17 formed stars at some point but only 13 of these retained them until $z=0$.  This leaves a total of 23 luminous subhalos at $z=0$, of which 13 have more than 50 baryon particles.

\begin{figure}
\centering
\resizebox{0.3\textheight}{!}{\includegraphics{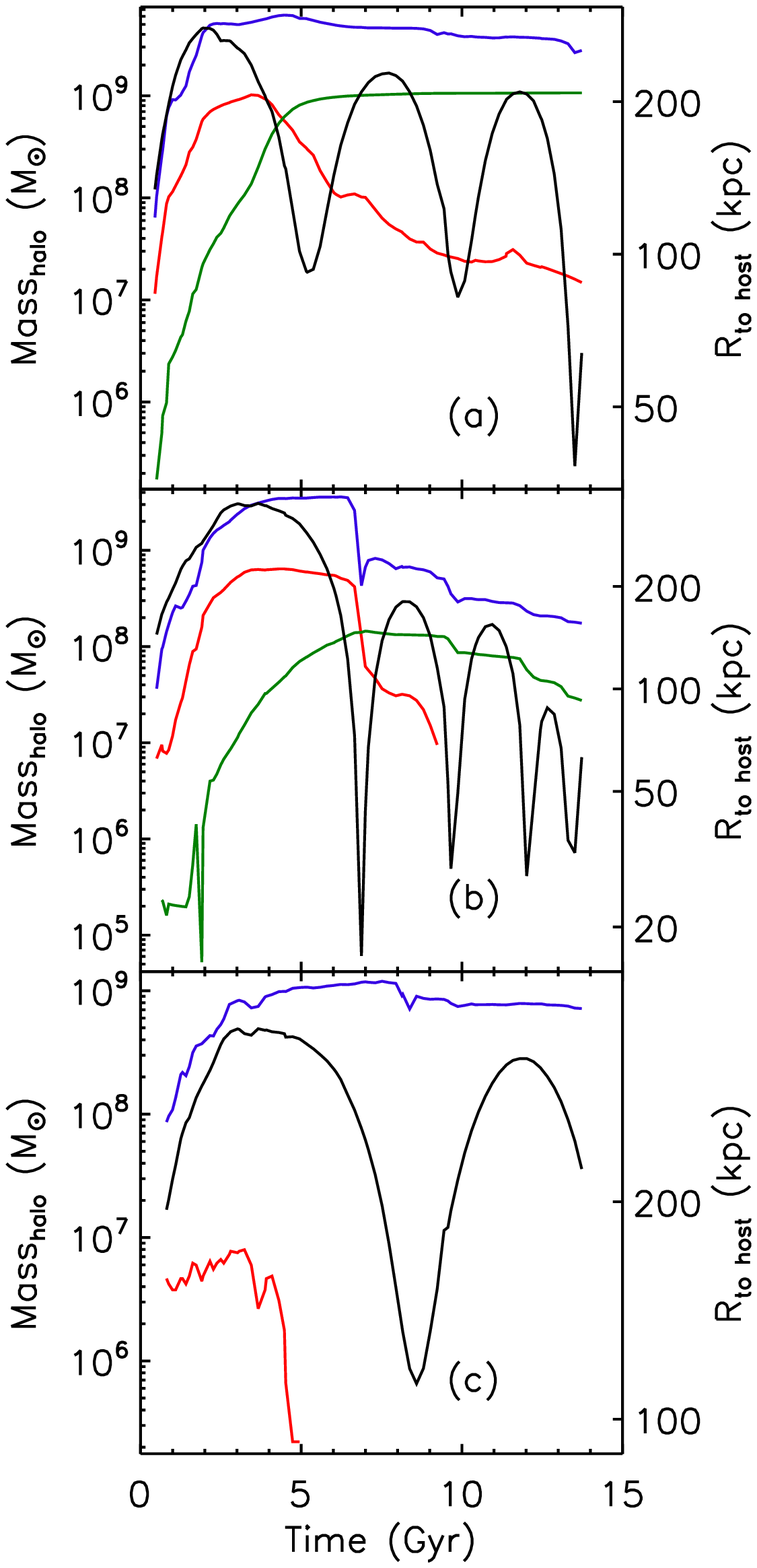}}
\caption{The time evolution of the distance to the host and the mass of different components of three subhalos. The black line corresponds to the distance to the host at each output on the right-hand axis on each plot. The coloured lines correspond to the masses on the left-hand axis: dark matter (blue), gas (red) and stars (green). The top subhalo (a) is massive enough to retain both gas and stars to $z=0$ (final mass: $3.9\times10^{9}$~M$_{\sun}$). The centre subhalo (b)  retains stars but not gas at $z=0$ (final mass: $2.0\times10^{8}$~M$_{\sun}$). The bottom subhalo (c) loses all its baryons before $z=0$, making it a dark satellite (final mass: $7.2\times10^{8}$~M$_{\sun}$) despite the fact that it ends up heavier than the luminous subhalo in the centre panel.}
\label{fig:timeevall}
\end{figure}

Figure \ref{fig:timeevall} shows examples of the time evolution of these subhalos' mass and distance to the host at each output, representing an upper limit on the subhalos' closest distance to the host. The top panel shows subhalo (a) that retains gas and stars at $z=0$, the middle panel shows subhalo (b) that retains stars but not gas at $z=0$, and the bottom panel (c) shows a dark satellite that has no baryon particles at $z=0$. Close passages to the center of the main halo most strongly removes gas and to a lesser extent, dark matter and stars through tidal stripping. Stars tend to sit at the centre of the subhalo and are less vulnerable to stripping than the dark matter around the subhalo's exterior.  That is, subhalos that form stars before they lose their gas retain those stars until $z=0$.  If stars are not formed before their first close passage then the subhalo will never form stars and becomes a dark satellite. The quantity of gas lost is also determined by the subhalo's proximity to the host. For example the luminous first and second subhalos (a) and (b) start out with similar mass, but the latter has a pericentre that is more than twice as close to the host and as a result, it loses all its gas by $z=0$. Central location is not the full story. The dark satellite (c) is farther from the host galaxy than either of the luminous ones and has a higher mass at $z=0$ than the subhalo (b), but it never forms stars and its mass in gas is much lower before being lost early in the simulation. There are more factors than a subhalo's mass at $z=0$ and its proximity to the host galaxy that decide if it is luminous at $z=0$, which we will explore in this section.

\subsection{Ultraviolet Background}
Quantifying how much gas a subhalo loses due to the UV background involves some degree of subjectivity. Instead of merely counting gas that was in the subhalo, it will involve determining how much gas was never in the subhalo but ought to have been.

We confine our analysis of the evolution of subhalos to the time after $z=10.8$, when the first halos get larger than the minimum group finder particle limit and are virialized. One of the earliest mechanisms that removes gas from subhalos is the reionising UV background radiation that is emitted from the first luminous objects.  This radiation makes its first impact at $z=9.9$ in these simulations.

Most of the subhalos cannot be identified until a few outputs after $z=9.9$. To enable the analysis of the effects of UV radiation, every dark matter particle within $r_{vir}$ is matched with a ``twin'' gas particle at the initial conditions. In our halo-by-halo analysis, dark matter twin particles are defined as subhalo members at the time their subhalo reaches its maximum mass. The evolution of the gas twins of these member particles is then traced from the earliest output onwards. The ensemble of the twin gas particles are referred to as the ``background gas'' later in this section.

\begin{figure}
\centering
\resizebox{0.45\textwidth}{!}{\includegraphics{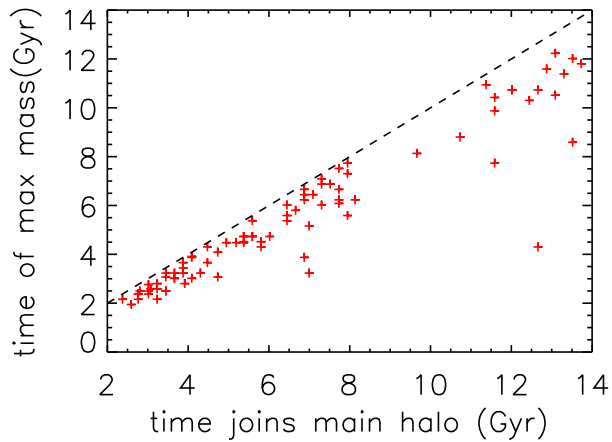}}
\caption{For all our subhalos, the time of maximum mass versus the time they join the main halo. Most of them reach their maximum mass shortly before becoming substructure. The dotted line marks where the time of maximum mass is equal to the substructure time, showing that none of the subhalos achieve their maximum mass after becoming substructure.}
\label{fig:tmmax_ss}
\end{figure}

The mass lost due to reionisation is defined as the gas that had a dark matter twin in a given subhalo, but itself was never contained in that subhalo. The reason that this gas was not in the subhalo is that subhalos are unable to contain gas with a temperature higher than the subhalo virial temperature ($T_{vir}$). Low mass subhalos will thus contain dark matter without its gas twin. Note that the twins are calculated from dark matter in the subhalo at the time of the subhalo's maximum mass and not reionisation. This is important because the way subhalo tracing works, only one subhalo is labelled as ``the'' subhalo at any given timestep. However, the larger subhalos reach their maximum mass as a result of the merger of several smaller subhalos. Determining the UV loss at a subhalo's maximum mass accounts for loss in each of the individual subhalos. Figure \ref{fig:tmmax_ss} shows that satellites reach their maximum mass prior to being captured by the main halo, typically immediately before.

Subhalos that did not virialize before reionisation generally do not contain gas. The subhalos that virialize after reionisation, but contain gas that goes on to form stars, do typically have gas without a dark matter twin in the subhalo. In the face of such a mismatch between gas and dark matter particle members we developed the following analysis to justify why the amount of background gas that ends up in the subhalo correlates to UV loss.

In order for a gas particle to have enough energy to overcome the potential of the subhalo and escape, its temperature must be greater than $T_{vir}$, where:
\begin{equation}
T_{vir}=\frac{2G\mu m_{p} M_{subhalo}}{3kR_{subhalo}}
\label{eqn:tvirial}
\end{equation}
where $G$ is the gravitational constant, $\mu$ is the mean molecular weight (where the typical value is 0.6 for ionised gas), $m_{p}$ is the proton mass, $k$ is Boltzmann's constant, and $M_{subhalo}$ is the subhalo's mass. $R_{subhalo}$ is the distance between the halo's centre of mass and the most distant member particle. 

\begin{figure}
\centering
\resizebox{0.3\textheight}{!}{\includegraphics{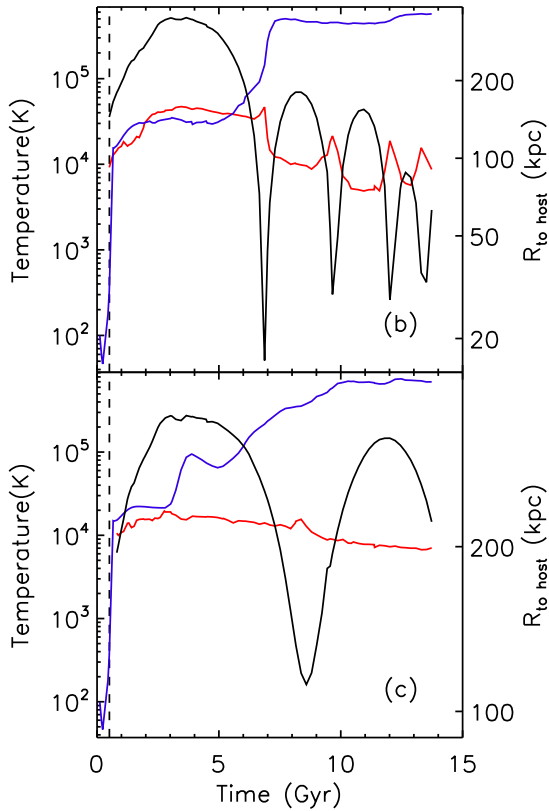}}
\caption{Evolution of background gas mean temperature of the background twin gas (blue) and $T_{vir}$ as given by Equation \ref{eqn:tvirial} (red) for subhalos (b) and (c) from Figure \ref{fig:timeevall}. Temperature corresponds to the left axis, and the subhalos' orbits (black, solid) to the right. The vertical dashed line corresponds to $z=9.9$ when the UV background is turned on. Even though the subhalo on the bottom is more massive at $z=0$, it is unable to retain as much gas as the lower mass subhalo on the top due to the fact that its background gas is consistently hotter than $T_{vir}$.}
\label{fig:tempev}
\end{figure}

Figure \ref{fig:tempev} shows the evolution of the background mean temperature (twin particles) and $T_{vir}$ in halos (b) and (c) from Figure \ref{fig:timeevall}. The subhalo in the top panel (b) retains its baryons through reionisation because of its higher mass and consequently higher T$_{vir}$.  While halo (c) is more massive at $z=0$ than (b), it is dark because the twin gas of its dark matter was hotter than $T_{vir}$. The subhalos that form earliest capture the most gas and form the most stars.

\begin{figure}
\centering
\resizebox{0.45\textwidth}{!}{\includegraphics{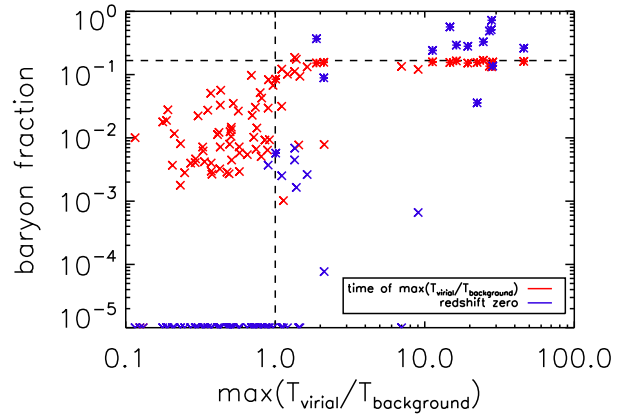}}
\caption{The baryon fraction versus the maximum ratio between the virial and background gas temperatures that a subhalo ever obtains over its lifetime. The horizontal dashed black line represents the cosmic mean, 0.17, while the vertical dashed black line separates the subhalos that achieved a $T_{vir}$ higher than the background gas temperature on the right from those that did not on the left. The asterisks represent subhalos that have ten or more baryon particles at $z=0$, while Xs represent subhalos that have fewer than ten baryon particles at $z=0$, corresponding to the resolution limit found in Figure \ref{fig:cumemag}. Red shows the subhalos' baryon fraction at the time of their maximum ratio of  $T_{vir}/T_{background}$, while blue shows the subhalos' baryon fraction at $z=0$. With one exception, every subhalo that is luminous at $z=0$ is to the right of the vertical line.}
\label{fig:bfractemp}
\end{figure}

Figure \ref{fig:bfractemp} shows how the ratio of $T_{vir}$ with the mean gas twin temperature effects the final baryon fraction. Specifically, we compare the maximum ratio of $T_{vir}$ to background gas mean temperature versus the baryon fraction at two times: $z=0$ and the time at which this temperature ratio is a maximum. The subhalos with T$_{vir}$/T$_{background} > 1$ at some point are more likely to be near the cosmic baryon fraction. With one exception, the subhalos that contain baryons at $z=0$ are a subset of these.  The one exceptional satellite started out near to the host and gathered enough gas early in its formation to form enough stars to remain luminous over its subsequent several close passages to the host, even though it lost all its gas on its second passage. Since none of the subhalos that never contained their twins held onto any of their baryons, it appears that the ``twin'' particle analysis is robust. Therefore, we can safely define gas lost due to UV background as the mass of twin gas particles that never entered subhalo.

\subsection{Ram Pressure Stripping}
One of the difficult aspects of this study is that the mass loss mechanisms involve the interaction between two gas phases with significantly different properties.  This circumstance is one where SPH struggles.  \citet{Agertz2007} showed that SPH has trouble modelling ram pressure stripping, particularly when the Kelvin-Helmholtz time is important.  However, \citet{Mayer2006} showed that SPH can model stripping when $\tau_{\mathrm{dyn}} < \tau_{\mathrm{KH}}$, and our g15784 satellites typically fall into this regime (\citealt{Mayer2006} show that $\tau_{\mathrm{KH}} \ga 4$~Gyr when they reach their minimum at pericentre, compared to satellite dynamical times of $\tau_{\mathrm{dyn}} \approx 0.2$--$6$~Gyr).

With these caveats in mind, we do a classical ram pressure analysis \citep{Gunn1972} to see how close these simulations come to reality.  In order to quantitatively measure the effect ram pressure stripping  has on our subhalos, we use the criterion from \citet{Grebel2003}:
\begin{equation}
P_{ram} \approx \rho_{hhg} v_{subhalo}^{2} > \frac{\sigma_{subhalo}^{2}\rho_{gas}}{3}
\label{eqn:pram}
\end{equation}
where $P_{ram}$ is the ram pressure, $\rho_{hhg}$ is the gas density in the hot halo gas around the subhalo, $v_{subhalo}$ is the subhalo's velocity relative to the host galaxy, $\sigma_{subhalo}$ is the velocity dispersion of the gas in the subhalo, $\rho_{gas}$ is the average density of the gas in the subhalo. Here the subhalo's velocity dispersion is defined as:
\begin{equation}
\sigma_{subhalo}^{2}=\frac{3}{5}\frac{GM_{subhalo}}{R_{subhalo}}
\label{eqn:veldisp}
\end{equation}
where $M_{subhalo}$ is the subhalo's mass, and $R_{subhalo}$ is the subhalo's radius.

\begin{figure}
\centering
\resizebox{0.45\textwidth}{!}{\includegraphics{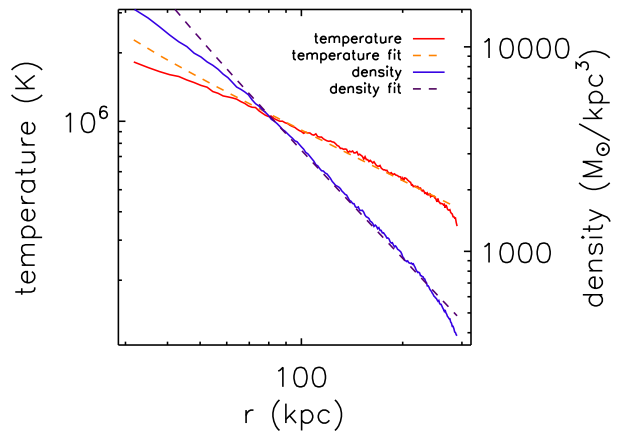}}
\caption{The gas temperature (red) and density (blue) profiles of g15784 at $z=0$ for the outer disk at $r \ge 30$kpc. Power laws fit to $\rho \propto r^{-1.7}$ (dashed purple) and $T \propto r^{-0.73}$ (dashed orange) are shown for reference.}
\label{fig:gden}
\end{figure}

The gas density of the hot halo gas around each halo is defined as the average density of the $n$ nearest gas particles, where $n$ is twice the number of particles in the subhalo to a maximum of 4000. The density and temperature structure of the gaseous halo of g15784, through which the subhalos pass, is shown in Figure~\ref{fig:gden} with substructure removed.

Every gas particle that leaves during the timesteps when the ram pressure exceeds the internal pressure of the halo is classified as leaving due to ram pressure stripping except for those particles that qualified for stellar feedback (\S\ref{sec:sf}).

\begin{figure}
\centering
\begin{tabular}{cc}
\resizebox{0.45\linewidth}{!}{\includegraphics{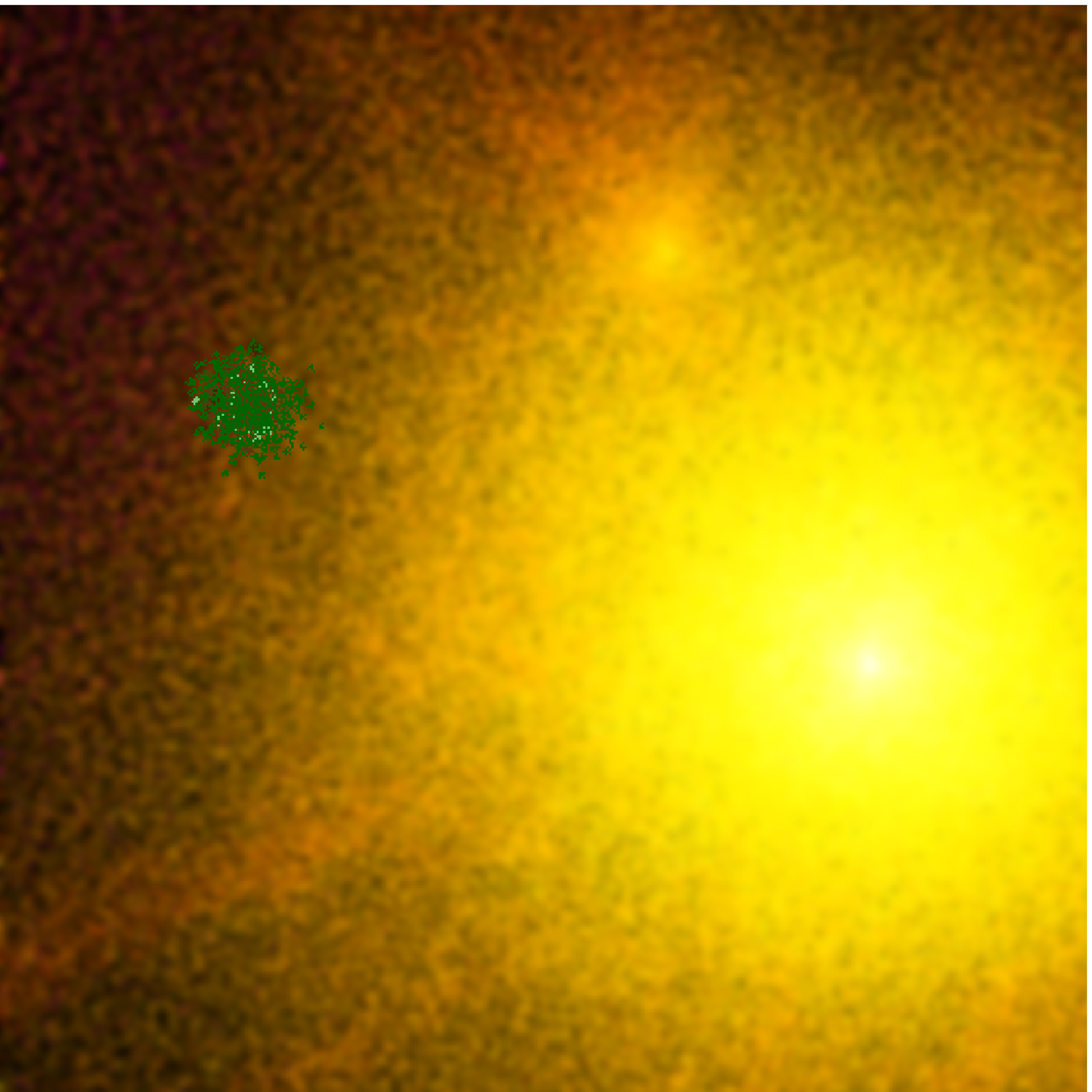}} &
\resizebox{0.45\linewidth}{!}{\includegraphics{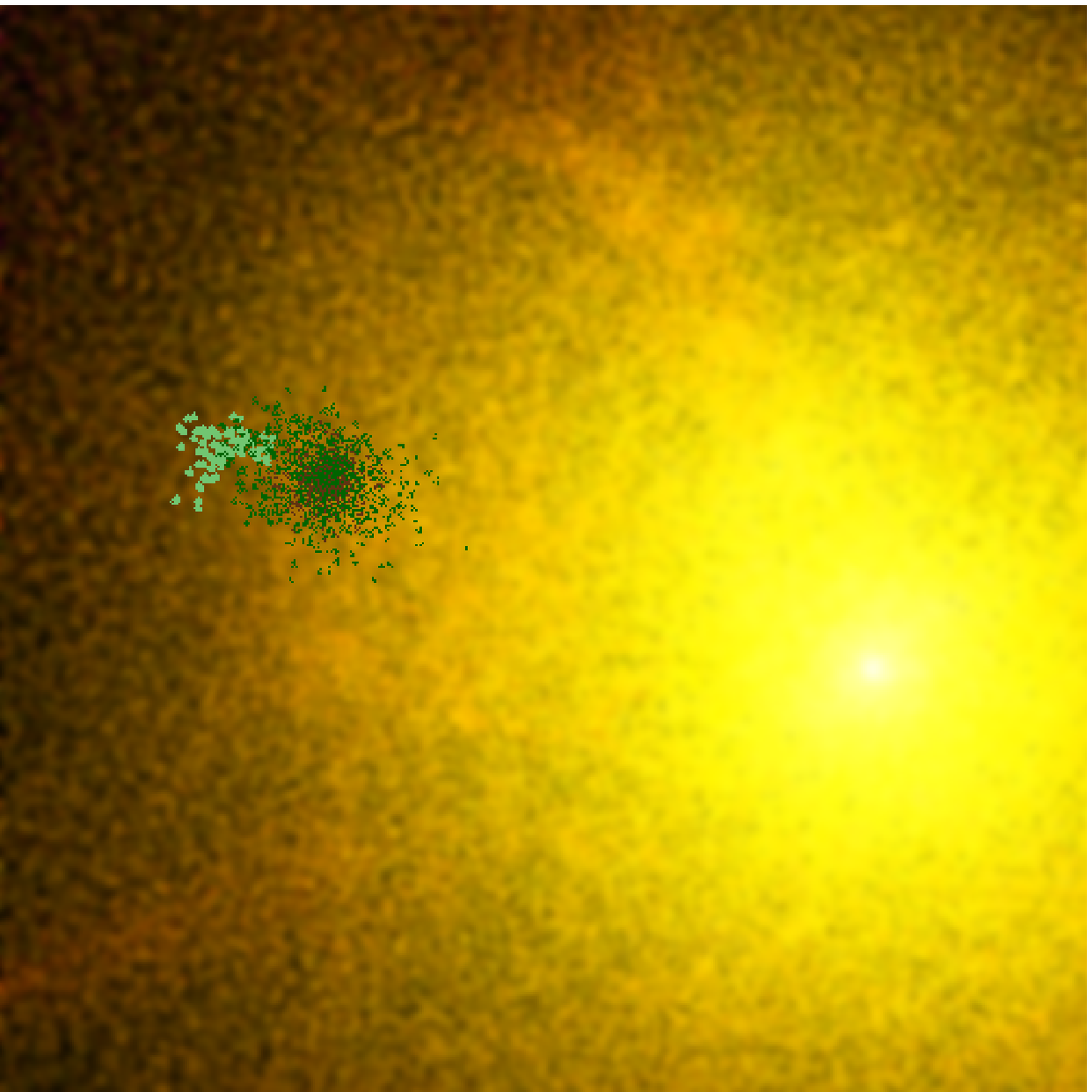}}
\end{tabular}
\caption{How ram pressure stripping removes gas from a $1.7\times10^{9}$~M$_{\sun}$ subhalo between $z=0.26$ (left) and $z=0.16$ (right). The dark matter (dark green), gas (light green) and stars (turquoise, none present here) are marked at the time of maximum mass, while the bottom layer (brown, covered the other layers) is the subhalo at the present output. Ram pressure stripping has affected this subhalo, removing all of the gas it pocessed at the time of maximum mass, but without strengthening the ram pressure stripping diagnosis by a factor of 10 the individual gas particles would not be marked as ram pressure stripped. The background colours denote the temperature of the surrounding gas, ranging from dark blue (colder, $10^{1.5}$~K) through to white (hotter, $10^{7}$~K).}
\label{fig:ram013}
\end{figure}

However, we also generated movies of twelve low-mass subhalos and visually inspected them to see if ram pressure or tides removed their gas. Figure \ref{fig:ram013} gives an example of a subhalo that loses its gas due to ram pressure. As a subhalo approaches the host galaxy, it enters the hot halo gas.  In low enough mass subhalos, its gas gets left behind, cleanly separating from its dark matter. Figure \ref{fig:ram013} shows that the gas maintains the shape of the subhalo for a several tens of Myrs. This contrasts with the signature of tidal stripping where tidal forces elongate the matter ahead and behind the satellite's orbit. Typically, ram pressure stripping occurs farther out from the host than tidal stripping.

\begin{figure}
\centering
\begin{tabular}{cc}
\resizebox{0.45\linewidth}{!}{\includegraphics{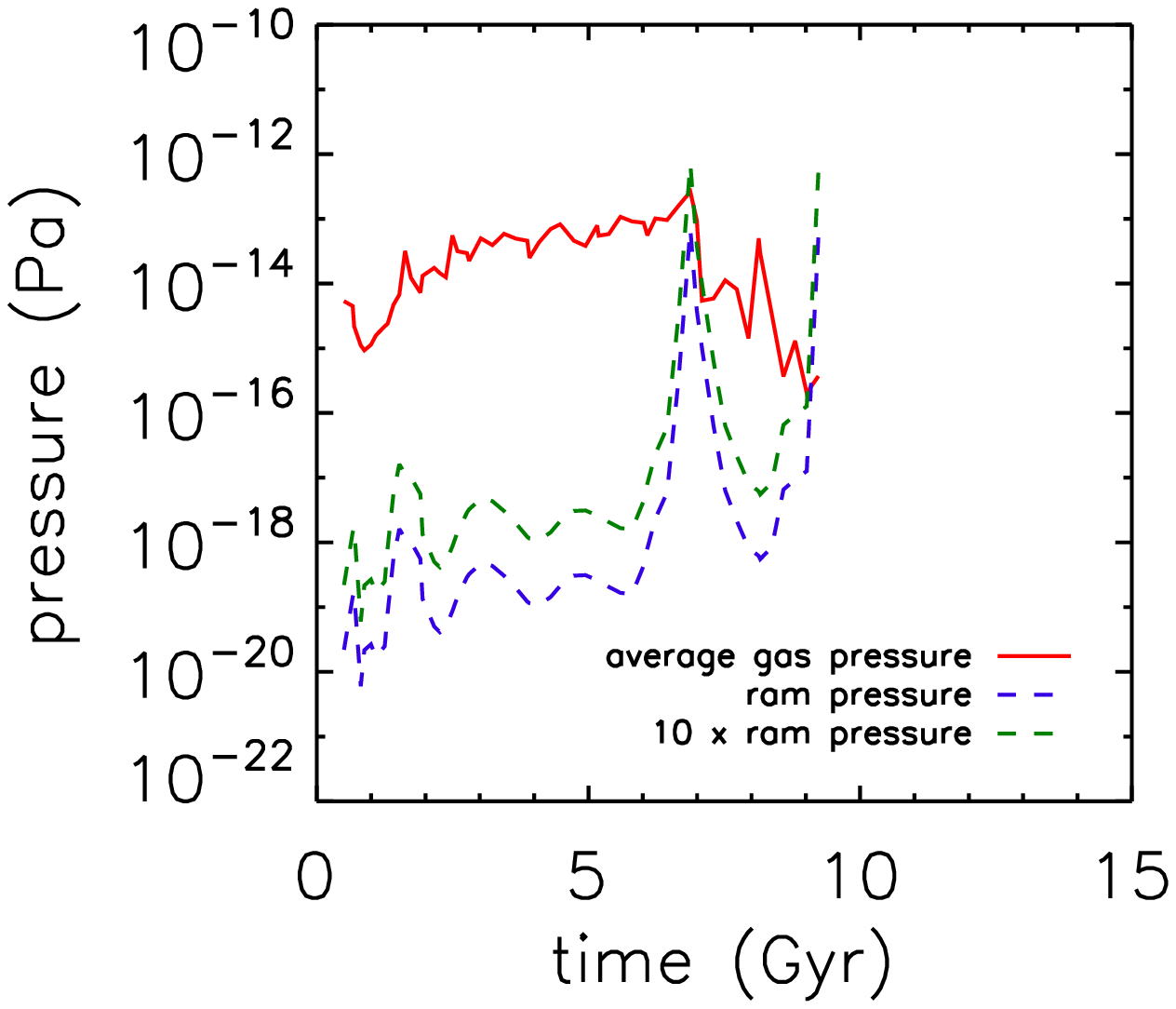}} &
\resizebox{0.45\linewidth}{!}{\includegraphics{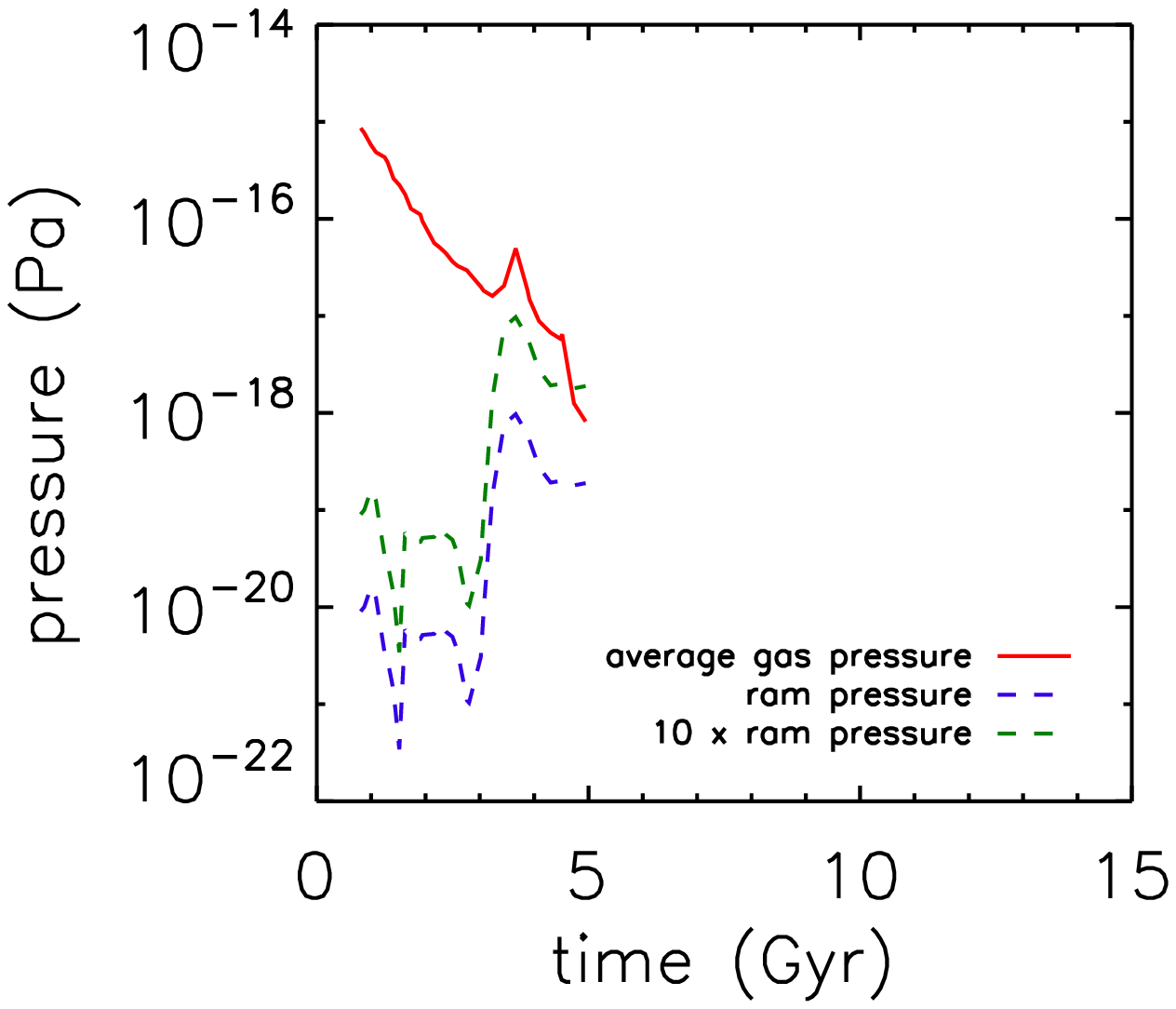}}
\end{tabular}
\caption{Pressure versus time for subhalos (b) and (c) from Figure \ref{fig:timeevall}. Internal pressure is given by the solid red line, ram pressure from the hot halo gas is the blue dashed line, and ram pressure strengthened by a factor of 10 is the green dashed line. Visual inspection of the subhalos revealed that the one in the left panel was affected both by ram pressure stripping and later tidal stripping, while the subhalo on the right was affected predominantly by ram pressure stripping.}
\label{fig:presev}
\end{figure}

Our definition for ram pressure underestimated the effect that is visible in the simulations. This seems to be a numerical effect and requires a multiplication of the calculated ram pressure by a factor of 10.  The factor of 10 may compensate for the use of the mean satellite gas pressure instead of the pressure in the outer regions where particles are getting removed. Figure \ref{fig:presev} shows how the subhalo's gas pressure and ram pressure evolves in two subhalos. The factor of 10 increases the ram pressure so that it is comparable with the subhalo's gas pressure. In the future, a comparison of higher-resolution simulations, as well as grid codes, to ours would be useful to see whether they exhibit results closer to the analytic determination.

\subsection{Tidal Stripping}
Next, we examine mass loss due to tidal forces using a simple spherical approximation for the host halo and the satellite. For each subhalo, the tidal force particles feel from the host is compared with the force they feel from their internal gravity. The subhalo's tidal radius is the place at which its self gravity is less than the tidal force of the host galaxy (e.g \citet{Hayashi2003}). Due to Newton's Theorem, assuming spherical symmetry, one need only consider the mass interior to a given particle.  The gravitational force on the particle from inside the particle's orbit is
\begin{equation}
F_{subhalo}=\frac{GM_{subhalo}(r)m_{particle}}{r^{2}}
\label{eqn:halograv}
\end{equation}
where $r$ is the distance from the particle to the centre of its subhalo, $m_{particle}$ is the mass of the particle, and $M_{subhalo}(r)$ is the subhalo's mass interior to $r$. The tidal force that the particle feels from the host, if the host galaxy is approximated by a point mass and assuming all its mass is contained inside the satellite, is the differential pull between the particle's position in the satellite and the satellite's centre:
\begin{equation}
\delta F_{tidal}=\frac{-2GM_{host}m_{particle}r}{R_{host}^{3}}
\label{eqn:halotide}
\end{equation}
where $R_{host}$ is the distance between the subhalo and the host, and $M_{host}$ is the mass of the host.
Therefore, the condition for when the particle feels a greater tidal force than gravitational from its own subhalo is given by:
\begin{equation}
\frac{M_{subhalo}(r)}{r^{3}}<\frac{2M_{host}}{R_{host}^{3}}
\label{eqn:tidecon}
\end{equation}
A particle that passes this test qualifies for tidal stripping. We emphasize that this a spherical approximation, given that the subhalos occasionally have their shape distorted, though the distortion is usually symetrical. We did try varying the strengh of the tidal force, as we had done for ram pressure stripping, but found that it did not change results much and so this method has some degree of robustness.

\begin{figure}
\centering
\begin{tabular}{cc}
\resizebox{0.45\linewidth}{!}{\includegraphics{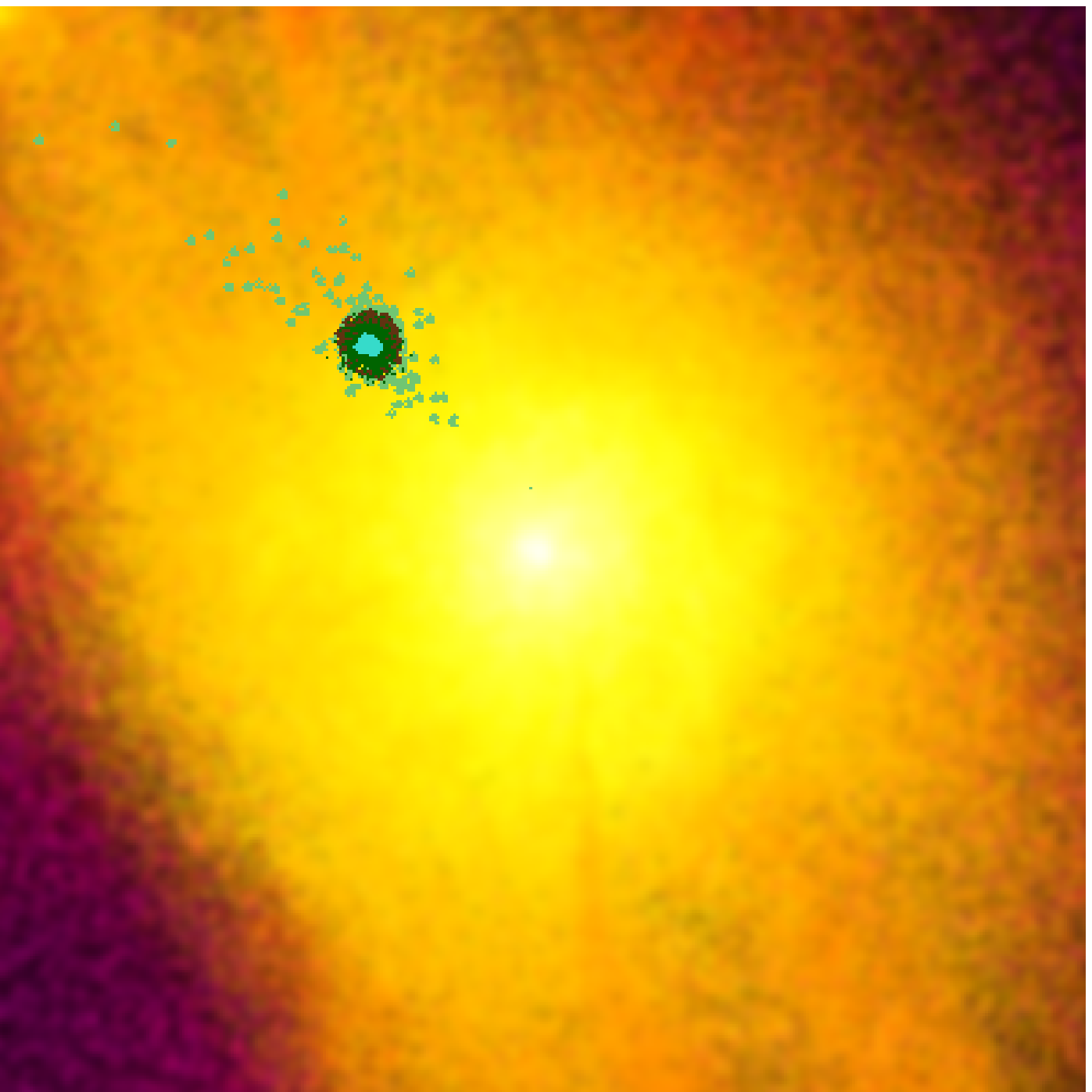}} &
\resizebox{0.45\linewidth}{!}{\includegraphics{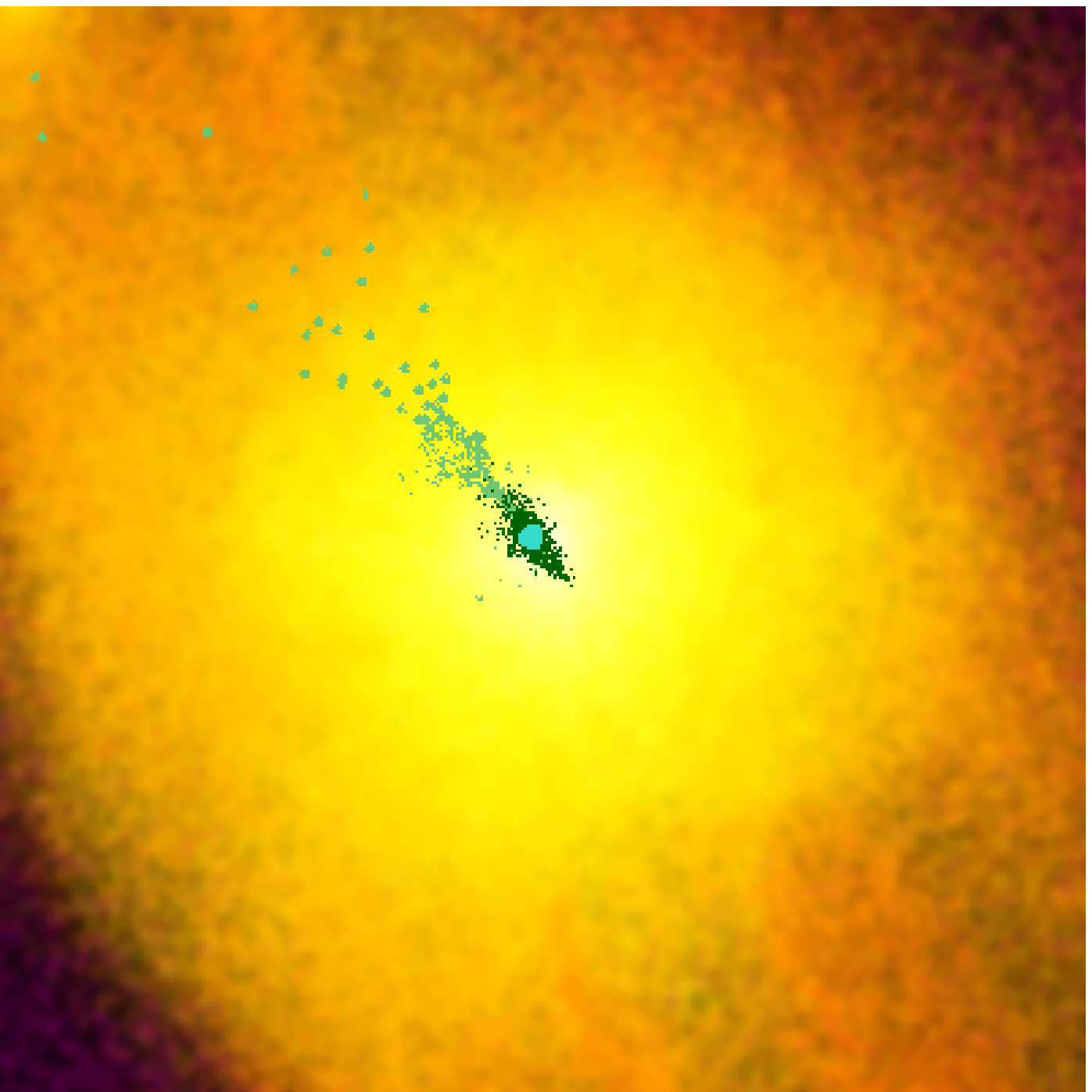}}
\end{tabular}
\caption{How tidal stripping removes gas from a $2.0\times10^{8}$~M$_{\sun}$ subhalos between $z=1.0$ (left) and $z=0.8$ (right). The dark matter (dark green), gas (light green) and stars (turquoise) are marked at the time of maximum mass, while the bottom layer (brown, covered the other layers) is the subhalo at the present output. Note how the tidal stripping has already begun at $z=1.0$, and by $z=0.8$ the tidal tail is prominent. The dark matter has been compressed, while the stars sit more safely in the subhalo's centre. The background colours are gas temperature as in Figure \ref{fig:ram013}.}
\label{fig:tide017}
\end{figure}

Figure \ref{fig:tide017} shows an example of a subhalo being tidally stripped of its gas.  The tidal force pulls material out in leading and trailing arms. Additionally, material is stripped off the outside first before the material on the inside.

Regarding the possibility of tidal stirring, Figure \ref{fig:timeevall} shows that there was not significant star formation following close passages of satellites.  During these close passages, dark matter is often stripped and a large fraction of gas is stripped.  It is possible that these simulations are too low of a resolution to model tidal stirring like was seen in \citet{Mayer2006}.

\subsection{Stellar Feedback}
\label{sec:sf}
Often, much credit for the removal of baryons is given to stellar feedback \citep{Dekel1986, MacLow1999, Dekel2003}.  Supernovae release large amounts of energy into the ISM, which can be sufficient to liberate gas from halo potential wells.  Others have argued that the coupling between the stellar feedback and the ISM is insufficient to remove a significant amount of gas from subhalos. Determining how much gas stellar feedback removed in the simulations proved to be a challenging task.

One signature of stellar feedback in these simulations is that a gas particle has its cooling turned off, which allows it to maintain its high temperature due to the stellar energy release.  When it gains sufficient kinetic energy, it can escape the gravitational potential of the subhalo. We found this method of tracking stellar feedback to be very limiting, however. Outputs were limited to approximately one every 200 to 100 Myr. Cooling is typically shut off for 50 Myrs, so between $1/2$ and $3/4$ of particles whose cooling was turned off would be missed by simply counting particles whose cooling was shut off during an output.

Another signature of stellar feedback is the release of metals into the surrounding interstellar medium. In our simulations, a star particle releases about $300 $ M$_{\sun}$ of metals from type II supernovae to the nearest 32 gas particles.  The ejection is smoothed so that gas closer to the star receives more metals than particles further away. A typical gas particle will receive a few M$_{\sun}$ in metals from a star particle.  Gas can also receive metals from other gas particles through diffusion.  These metal transfers are typically $<1$ M$_{\sun}$. 

So, when gas had an increase in metals of $\ge 5$ M$_{\sun}$, it is likely that it was in the neighborhood of stellar feedback and we classify it as having been lost from the halo due to stellar feedback.

This is a conservative estimate because there may also have been cases where gas directly heated by stellar feedback acquired sufficient pressure to push out different gas, a process called ``mass-loading''. This is common in dwarf galaxies \citep{Vacchia2008}.  Stellar feedback could have augmented another mechanism like ram pressure stripping and gotten unlabelled as such.

\subsection{A Combination of Mechanisms}
\label{sec:combomech}

Using all the techniques described above, we now present a summary of which processes dominated accretion and loss of gas.

One confounding effect happened when massive subhalos pass through the pericentre of their orbit. Subhalos temporarily accreted a small quantity of gas and quickly lose it, possibly a numerical effect.  Because of this, we did not count mass loss of particles that entered and left subhalos after they reached their maximum mass.

Some particles left the subhalo more than once. In order to avoid double counting, only the last method by which a particle entered or left its subhalo was counted. We noted which gas particles were converted into stars, so that we could identify what portion of the gas mass decrement was due to star formation and what portion was due to gas leaving the subhalo.  The loss of many particles could not be classified, so mass loss is often classified as ``other'', emphasizing the schematic nature of this method.

\begin{figure}
\centering
\resizebox{0.45\textwidth}{!}{\includegraphics{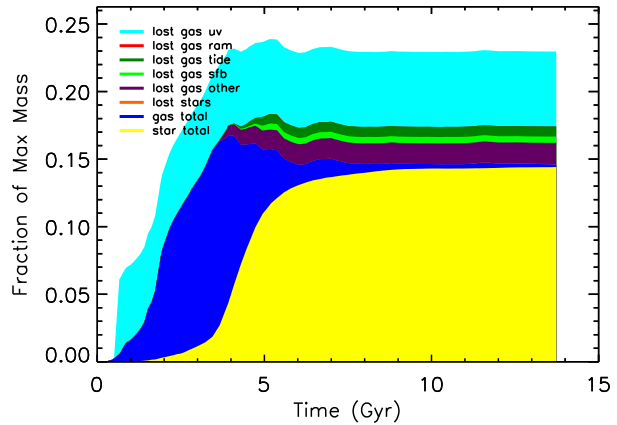}}
\caption{Mechanisms for baryon loss over time for the massive subhalo (a) (maximum mass: $7.4\times10^{9}$~M$_{\sun}$, final mass: $3.9\times10^{9}$~M$_{\sun}$) from Figure \ref{fig:timeevall}. Turquoise is the cumulative gas lost due to the UV background, red is the cumulative gas lost due to ram pressure stripping, dark green is the cumulative gas lost due to tidal stripping, light green the is the cumulative gas lost due to stellar feedback, purple is the cumulative gas lost due to undetermined causes, orange is the cumulative lost stars, blue is the gas at the current timestep, and yellow is the stars at the current timestep.}
\label{fig:mlosptcla}
\end{figure}

\begin{figure}
\centering
\resizebox{0.45\textwidth}{!}{\includegraphics{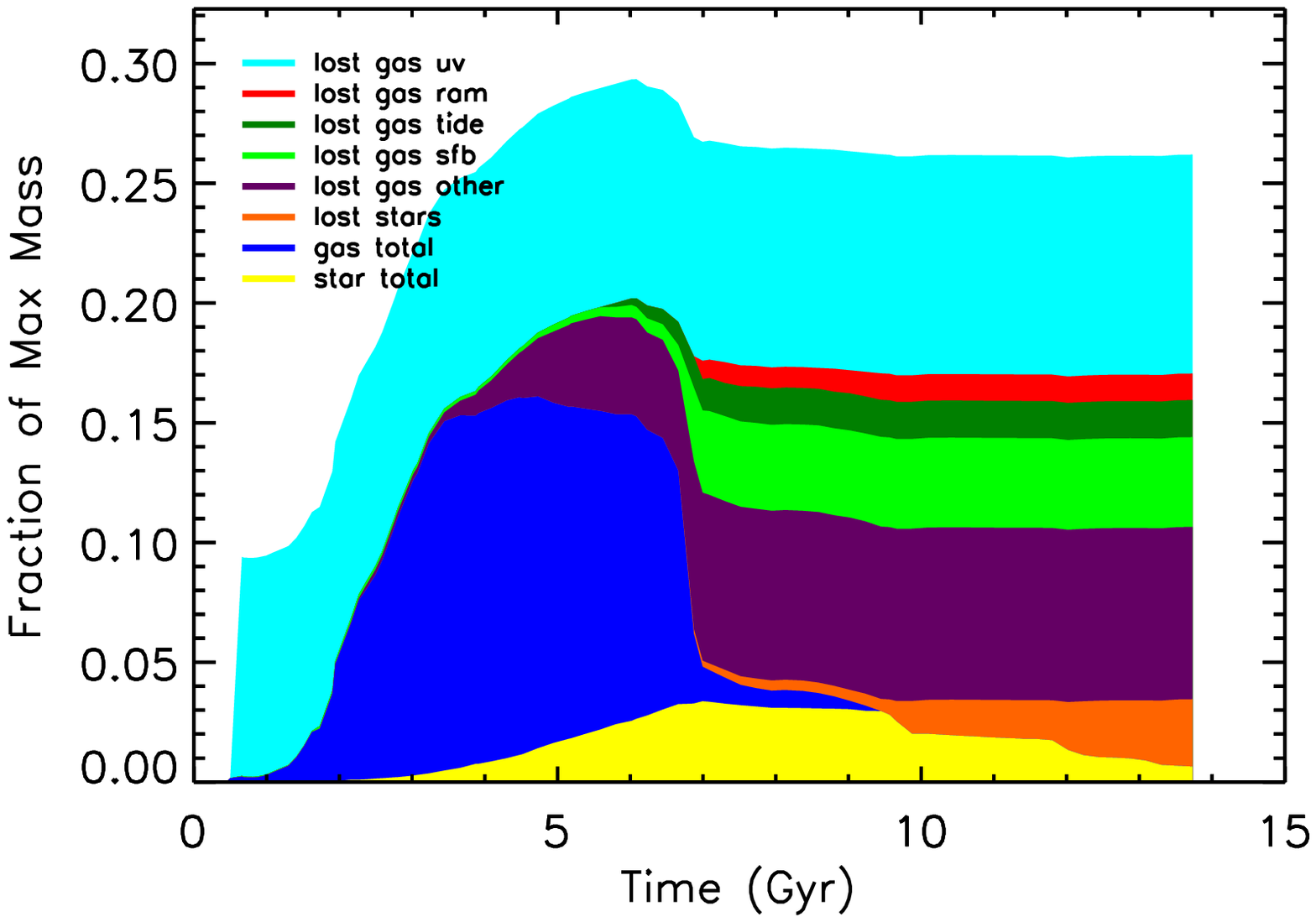}}
\caption{Mechanisms for baryon loss over time for the medium mass subhalo (b) (maximum mass: $4.2\times10^{9}$~M$_{\sun}$, final mass: $2.0\times10^{8}$~M$_{\sun}$) from Figure \ref{fig:timeevall}. The colour scheme follows Figure \ref{fig:mlosptcla}.}
\label{fig:mlosptclb}
\end{figure}

\begin{figure}
\centering
\resizebox{0.45\textwidth}{!}{\includegraphics{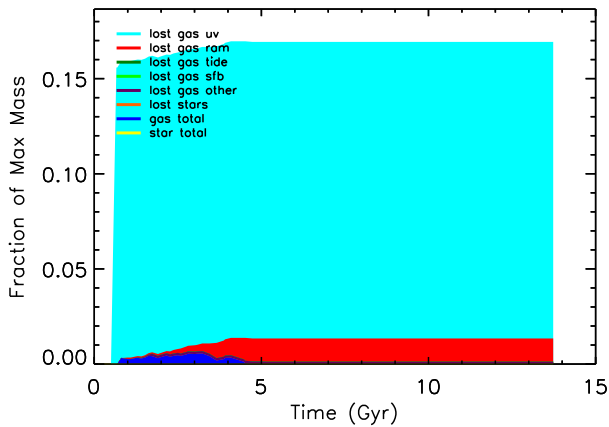}}
\caption{Mechanisms for baryon loss over time for the medium mass subhalo (c) (maximum mass: $1.4\times10^{9}$~M$_{\sun}$, final mass: $7.2\times10^{8}$~M$_{\sun}$) from Figure \ref{fig:timeevall}. The colour scheme follows Figure \ref{fig:mlosptcla}.}
\label{fig:mlosptclc}
\end{figure}

Figures \ref{fig:mlosptcla} to \ref{fig:mlosptclc} show the evolutionary history of the same subhalos as shown in Figure \ref{fig:timeevall}. These histories are coloured to indicate the amount of gas lost due to each of the mechanisms described above, as a fraction of each subhalo's maximum mass. Most of the baryons for the most massive subhalo (a) turn into stars quickly, and, other than the initial UV ionisation which prevents a significant amount of gas from being captured, any gas that is lost is usually due to tidal stripping or stellar feedback. After UV ionisation the medium mass subhalo (b) that retains its stars loses its gas largely due to stellar feedback with smaller contributions by tidal and ram pressure stripping, though there is a large number of unclassified particles as well. The medium mass subhalo (c) that ends up as a dark satellite lost almost all of its gas due to the UV ionisation, with the small remainder being stripped by ram pressure.

\begin{figure}
\centering
\resizebox{0.45\textwidth}{!}{\includegraphics{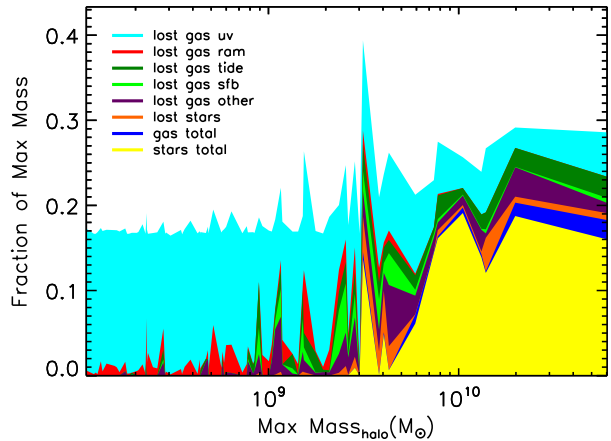}}
\caption{Mechanisms for baryon loss at $z=0$ for all subhalos, as a function of the maximum mass that a subhalo was able to achieve. The colour scheme follows Figure \ref{fig:mlosptcla}. The various loss mechanisms are cumulative over time, while the total gas and stars are for the current time. All values are a fraction of the subhalo's maximum mass. The luminous subhalos show an excess of baryons over their lifetime, above the cosmic mean of 0.17. The cutoff between luminous and dark satellites happens at about $2.0\times10^{9}$~M$_{\sun}$.}
\label{fig:mlosptclall}
\end{figure}

Figure \ref{fig:mlosptclall} combines all plots of type Figures \ref{fig:mlosptcla} to \ref{fig:mlosptclc} for all subhalos and shows as a fraction of each subhalo's maximum mass the gas lost due to each mechanism at $z=0$.  These mass loss fractions are plotted as a function of maximum mass rather than final mass because the sequence of mechanisms appears more clearly (as shown in Figure \ref{fig:msmtot}).  Figure \ref{fig:mlosptclall} shows that the massive subhalo (a) in Figure \ref{fig:mlosptcla} is no aberration.  It is common for the most massive subhalos to efficiently form stars and for tidal stripping to play the most important role in their mass loss. Medium mass subhalos tend to be more dominated by stellar feedback, since they are massive enough to form stars but light enough that they are more susceptible to losing their gas. Following Figure \ref{fig:mlosptclc}, lower mass subhalos lose significant mass due to UV reionisation and then much of the remaining mass is stripped by ram pressure.  

A dichotomy of evolutionary scenarios appears in Figure \ref{fig:mlosptclall}. Halos less massive than $\approx2.0\times10^{9}$~M$_{\sun}$ lose their gas due mostly to UV reionisation. Higher mass halos formed stars, lost less mass to reionization, and lost gas due to a variety of other mechanisms. This distinction disappears when the satellites are classified by their final mass as in Figure \ref{fig:msmtot} where the populations of baryonless and luminous subhalos overlap in terms of total mass at $z=0$.

Table \ref{tab:mechpercent} summarizes the results of Figure \ref{fig:mlosptclall} by dividing the subhalos into three categories (ones with gas and stars at $z=0$, ones without gas at $z=0$ but had stars, and ones that never formed stars and have no gas at $z=0$). UV ionisation is the most prominent for all subhalos. For the most massive subhalos tidal stripping followed, while stellar feedback and ram pressure stipping had little impact. For the subhalos massive enough to form stars at some point but which did not retain gas at $z=0$, after UV ionisation, stellar feedback was the most prominet mechanism, while tidal and ram pressure stripping were close in magnitude. Finally, UV ionisation was the most important for the subhalos that never formed stars, with some impact from ram pressure stripping. Tidal stripping and stellar feedback were negligible. Note the caveat: since it is impossible in all cases to clearly distinguish the mechanism that leads to the loss of a gas particle, the boundaries between the mechanisms are not clearly defined and the percentages should, therefore, be take as indicative of the relative importance of the various gas loss mechanisms.

\begin{table}
\centering
\begin{tabular}{l|lll}
\hline
At $z=0$: & no gas and & no gas but     & gas \\ 
          & never stars & have/had stars & and stars \\ \hline \hline
Min Mass (M$_{\sun}$)&  $1.10\times10^{8}$ & $7.79\times10^{8}$ &  $3.14\times10^{9}$ \\
Max Mass (M$_{\sun}$)&  $2.09\times10^{9}$ & $4.30\times10^{9}$  &  $6.03\times10^{10}$ \\ \hline
UV (\%)         &  15.73 &  11.19 &   6.33 \\
 Ram (\%) &   1.10 &   1.54 &   0.26 \\
 Tides (\%)    &   0.12 &   1.87 &   1.93 \\
 Sfb (\%)  &   0.01 &   2.08 &   0.64 \\
 \hline
\end{tabular}
\caption{The cumulative effect of each gas-loss mechanism (UV ionisation, ram pressure stripping, tidal stripping, and stellar feedback) given as an approximate percentage of the subhalos' maximum mass, averaged over all the subhalos in each category. The first category are the subhalos that had no gas at $z=0$ and never formed stars (represented by subhalo (c)), the second are those that have no gas at $z=0$ but had stars at some point in their history (represented by subhalo (b)), and the last category are those subhalos that retain both gas and stars at $z=0$ (represented by subhalo (a)). Also included are the minimum and maximum masses of the subhalos in each category.}
\label{tab:mechpercent}
\end{table}

\begin{figure}
\centering
\resizebox{0.45\textwidth}{!}{\includegraphics{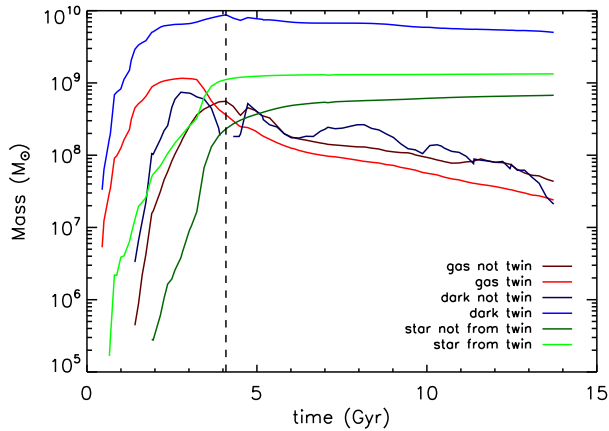}}
\caption{Time evolution of a  $7.1\times10^{9}$~M$_{\sun}$ mass subhalo's matter, broken down into dark matter present (twin) and not present (not twin) at the time of maximum mass, gas twinned or not from the maximum mass dark matter, and stars that did and did not come from twinned gas. The dashed vertical line is the time of maximum mass. Despite the difference in magnitudes between the twin and non-twin dark matter, the twin and non-twin gas particles are comparable, suggesting that the dark matter draws on gas outside its region of origin.}
\label{fig:twintrack}
\end{figure}

For the lower mass subhalos, the amount of mass lost adds up nearly to the cosmic baryon fraction ($\approx 0.17$), in part because our analysis relied on pairing dark and gas particles. However, not only do the higher mass subhalos contain more baryons than the cosmic fraction, they contain more \emph{stars} than the cosmic fraction. To understand how the higher mass subhalos form so many stars, we investigated the origin of these stars. Figure \ref{fig:twintrack} shows the mass evolution of a $7.1\times10^{9}$~M$_{\sun}$ subhalo that ends up with more than the cosmic baryon fraction in stars.  The mass evolution is divided into categories based on whether the particles were twins of the dark matter present at the maximum mass. While most of the stars formed from gas that was a twin of this dark matter, almost 10\% of the stars formed from gas that were twins of dark matter that were not members of this halo at its time of maximum mass, or any of the outputs immediately before and after the time of maximum mass.

\begin{figure}
\centering
\begin{tabular}{cc}
\resizebox{0.45\linewidth}{!}{\includegraphics{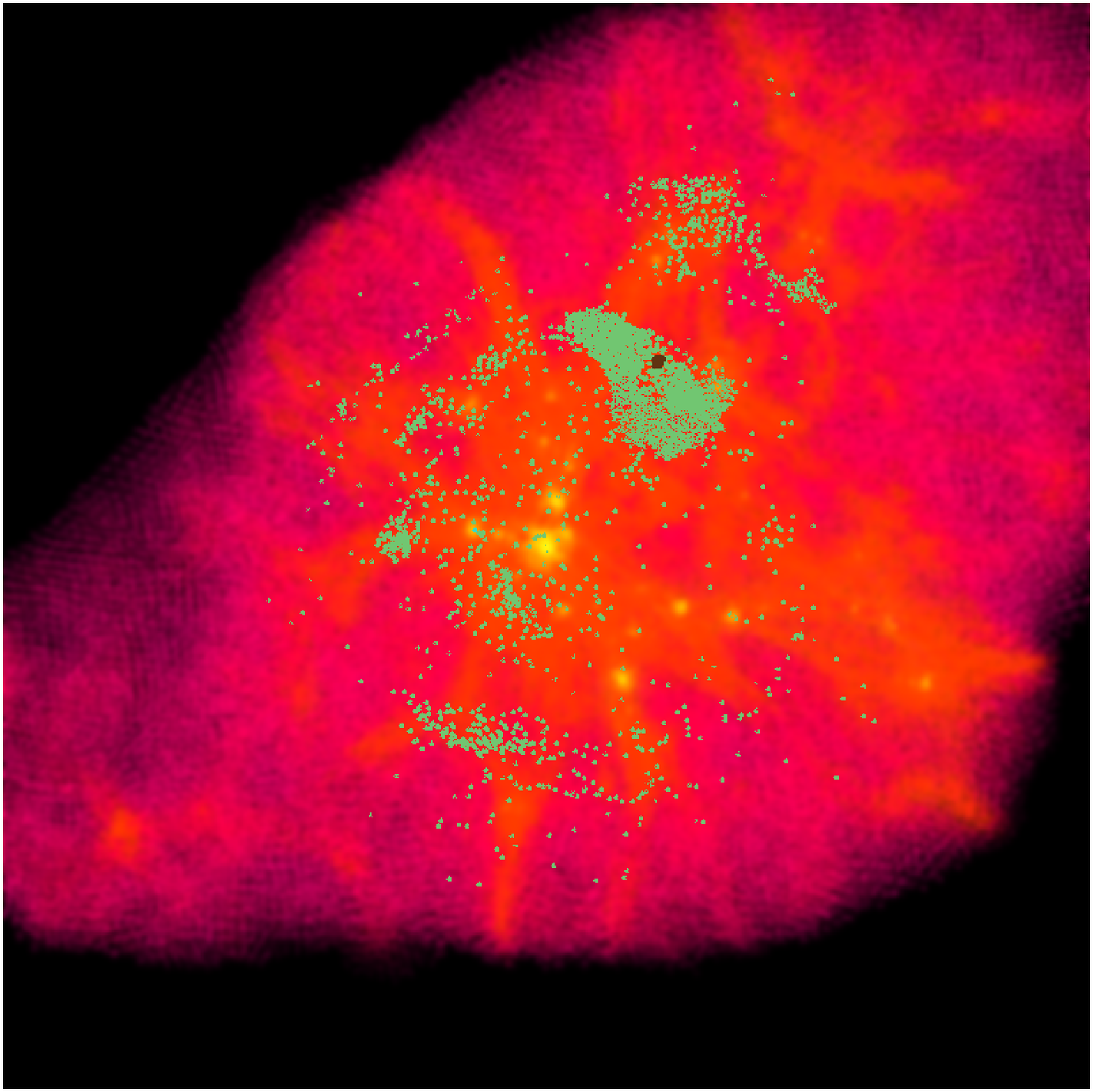}} &
\resizebox{0.45\linewidth}{!}{\includegraphics{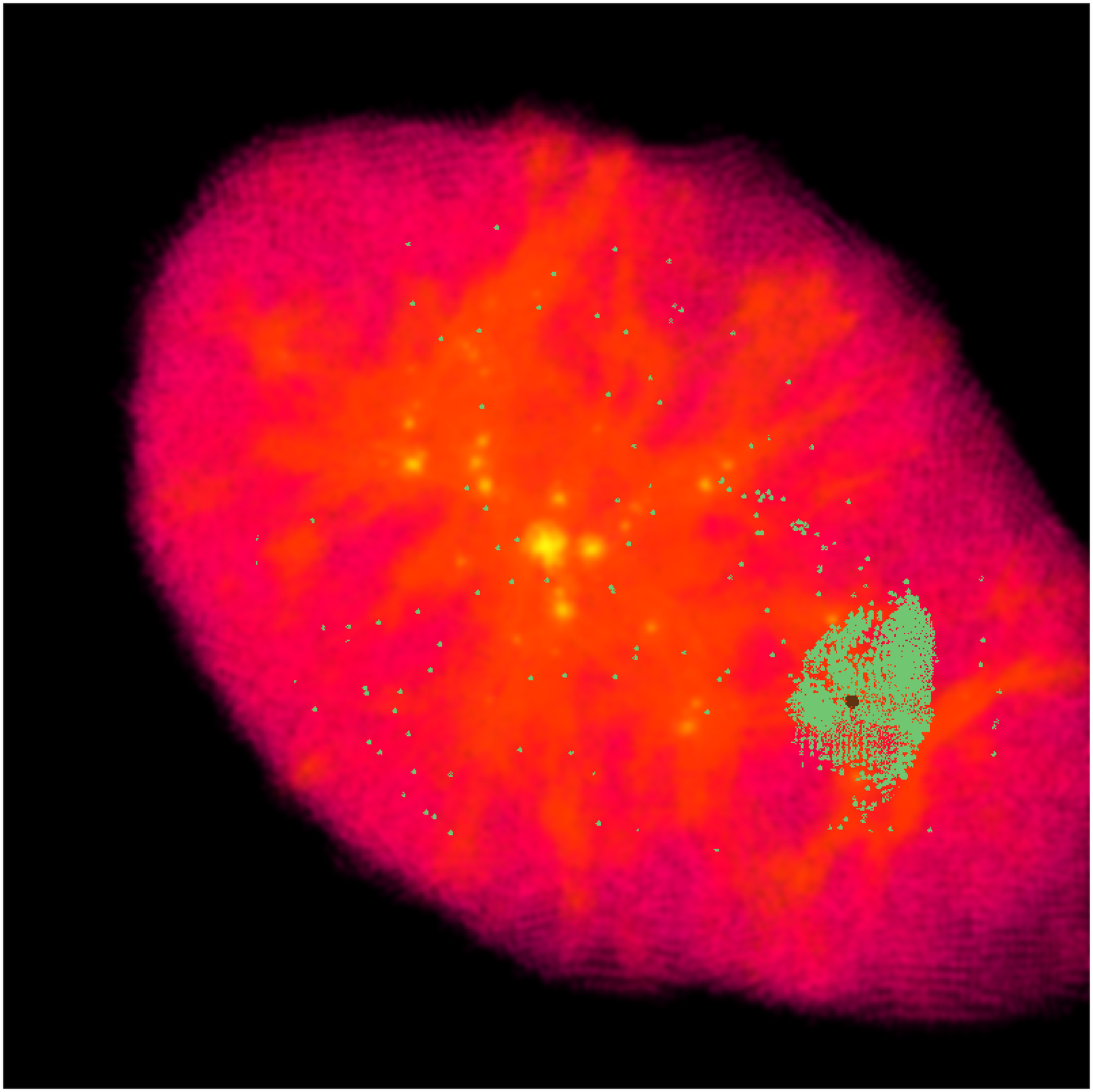}}
\end{tabular}
\caption{Snapshots at $z=6$ of a $7.1\times10^{9}$~M$_{\sun}$ mass subhalo (left) and a  $2.4\times10^{9}$~M$_{\sun}$ mass subhalo (right). The subhalos are in brown, while light green is the gas that will produce all the non-twin stars that will end up in the subhalos. This gas will eventually converge into the subhalos. The subhalo on the left will end up in a tight orbit around the host, and hence the several shells of gas that it will draw upon.  The background colours are gas temperature as in Figure~\ref{fig:ram013}.}
\label{fig:snotgorg}
\end{figure}

Figure \ref{fig:snotgorg} shows how this extra gas (marked as light green) comes from a much wider region than the dark matter (marked as brown).  While such accretion could be a numerical artifact of over-efficient gas cooling, it could also be a unique feature of satellites that orbit in high-density regions like a massive galaxy's hot halo.  The mechanism that appears in the simulations is that high mass satellites quickly form stars out of gas that are twins of member dark matter particle.  After they form stars and the stellar feedback cools down, there is less gas to provide pressure support to keep hot gas from the main halo out of the satellite.  So this gas is accreted, cooled, and finally forms stars.

\section{Conclusions and Discussion}
\label{sec:conclu}

To gain insight into the missing satellites problem, we compared the satellite luminosity functions of two simulated galaxies from the MUGS project (g5664 and g15784) with late-type galaxies.  

The cumulative number of luminous satellites in g15784 was only slightly higher than that observed in the Milky Way, though there were an excess of high luminosity satellites that created a ``knee'' in the satellite luminosity function that is not observed. Other SPH simulations of similar or lower resolution to ours have found that the missing satellites problem is no longer a matter of an order magnitude difference between the Local Group and simulated subhalo populations. When we compared our luminosity function of g15784 to a simulation of the same galaxy at a lower gas resolution both luminosity functions were similar in their area of overlap down to $M_V \approx -8.2$, though the low resolution run did suffer from having fewer gas particles than needed to properly resolve star formation and therefore it had fewer stars. A couple of our dwarfs had luminosities comparable to the recently discovered ultra-faint dwarfs, but at the resolution of these simulations, they had only one or two star particles, so it is impossible to draw any conclusions about the formation of fainter dwarf galaxies from these simulations. 

The satellite mass function of g15784 revealed a large population of dark satellites. In our more massive galaxy g15784 ($1.4\times10^{12}$~M$_{\sun}$) the subhalos constituted 6.0\% of the host galaxy's mass, while g5664's subhalos were 4.4\% of the host galaxy's mass. This fits within the range that \citet{Dalal2002} found from probing substructure with gravitational lensing, between 0.6\% and 7.0\%.  

We used two methods to determine how the dark satellites lost their baryons and the effect of negative feedback on the luminous satellites.

One method was to simulate the less massive galaxy than the Milky Way, g5664, using several different physical treatments.  The simplest included no UV or stellar feedback.  In the second, UV was added.  These two simulations were compared with the standard MUGS simulation that included both UV and stellar feedback.  The effect they had on the subhalo populations was significant.  UV feedback alone stopped star formation in all the satellites with total masses less than $2\times10^{9}$~M$_{\sun}$.  When stellar feedback was added, it reduced the luminosity of several additional subhalos so that only a couple of star particles formed in those subhalos before the feedback ejected all the remaining gas from the subhalos and eliminated the possibility of future star formation.  In more massive subhalos, the stellar feedback had little impact on reducing the star formation efficiency.  This unbalanced influence of the stellar feedback may have been due in part to the quantized feedback that was used in the MUGS simulations.

The second method was to analyse the individual evolution of satellites in a more massive halo.  We made a comprehensive study of the mechanisms that remove matter from subhalos by defining criteria for mass loss due to the UV background, tidal stripping, ram pressure stripping, and stellar feedback.  This analysis reiterated the strong impact ionisation had on low mass satellites.  We used metals to track stellar feedback, and found that its impact was largest on subhalos of medium mass that had formed stars, with lesser impact on the highest mass subhalos, and no impact on the lower mass subhalos.

A strange phenomenon was apparent in the higher mass subhalos.  These subhalos contained a higher fraction of their maximum mass in stars than the cosmic baryon fraction.  Subsequent analysis showed that accretion of baryons was not confined to the same limited region from which dark matter was accreted, but from a larger region surrounding the subhalo and even from across the hot gaseous host halo as the subhalo moved through its orbit.  While this may partially be another symptom of overcooling that has been long noted in simulations, it may also point to the enhanced baryonic accretion possible by subhalos in high density regions.

Stripping, either ram pressure or tidal, also plays a vital role in shaping the satellites that were analysed.  Ram pressure removed whatever gas remained in small subhalos that had most of their gas removed during ionisation.  In some cases, more gas was stripped from subhalos than would have been predicted based on a \citet{Gunn1972} analysis.  Tidal stripping removed most of the gas from the more massive satellites, making them comparable with Local Group dSphs rather than dIrrs.  The stripping became apparent once the subhalos crossed inside the virial radius.

Tidal stripping was important for the subhalos that had the closest encounters with the main galaxy.  In some cases, tidal stripping removed enough of the outer layers of dark matter that the total mass of the satellites dropped below the ionisation mass limit of $2\times10^{9}$~M$_{\sun}$. Because tidal stripping reduces the total mass of subhalos by different amounts, it is critical to organize the satellites by their maximum mass rather than their mass at $z=0$ to see a continuous behavior in the baryon fraction as a function of mass.  Many authors have noted the similarity in mass inferred in Local Group dSphs \citep{Bullock2000,Strigari2007,Pen2008} by extrapolating satellite total masses using NFW density profiles.  Our simulations point out that because of tidal stripping, these satellites may no longer contain that much mass.  However, those extrapolated masses may be similar to the maximum mass of the satellite, and the constant lower mass limit suggests a mass-dependent gas removal mechanism like ionisation.  What is not clear in the simulations is why the efficiency of star formation varies so much from dSph down to ultra faint galaxies if they did form from subhalos that were all the same mass.

We should also extend this work to include all MUGS galaxies to determine if there are specific factors in the environment or history of individual galaxies that affect the satellite population, as well as compare the subhalo populations outside of the main halos to the ones that end up within the virial radius of their main halos.

\section*{Acknowledgements}
We thank SHARCNET for generously providing supercomputers without which the MUGS galaxies would not be possible, NSERC for funding, and the referee for helpful comments. HMPC thanks the Canadian Institute for Advanced Research for support. GSS was a CITA National Fellow during part of this study and a Jeremiah Horrocks fellow during the rest of the study.

\bibliographystyle{mn2e}
\bibliography{references}

\clearpage

 \end{document}